\documentclass[preprint,12pt]{elsarticle}

\usepackage{packages}

\journal{arxiv.org}

\begin{document}

\begin{frontmatter}



\title{Robust Hole-Detection in Triangular Meshes Irrespective of the Presence of Singular Vertices}


\author[inst1]{Mauhing Yip}

\affiliation[inst1]{organization={Department of Engineering Cybernetics, NTNU},
            addressline={O. S. Bragstads Plass 2D}, 
            city={Trondheim},
            postcode={7034}, 
            country={Norway}}

\author[inst1]{Annette Stahl}
\author[inst2]{Christian Schellewald}

\affiliation[inst2]{organization={SINTEF Ocean},
            addressline={Brattørkaia 17c}, 
            city={Trondheim},
            postcode={7010}, ,
            country={Norway}}

\begin{abstract}
In this work, we present a boundary and hole detection approach that traverses all the boundaries of an edge-manifold triangular mesh, irrespectively of the presence of singular vertices, and subsequently determines and labels all holes of the mesh. 
The proposed automated hole-detection method is valuable to the computer-aided design (CAD) community as all half-edges within the mesh are utilized and for each half-edge the algorithm guarantees both the existence and the uniqueness of the boundary associated to it. As existing hole-detection approaches assume that singular vertices are absent or may require mesh modification, these methods are ill-equipped to detect boundaries/holes in real-world meshes that contain singular vertices. 
We demonstrate the method in an underwater autonomous robotic application, exploiting surface reconstruction methods based on point cloud data. In such a scenario the determined holes can be interpreted as information gaps, enabling timely corrective action during the data acquisition. However, the scope of our method is not confined to these two sectors alone; it is versatile enough to be applied on any edge-manifold triangle mesh.
An evaluation of the method is performed on both synthetic and real-world data (including a triangle mesh from a point cloud obtained by a multibeam sonar). The source code of our reference implementation is available: {\color{red} https://github.com/Mauhing/hole-detection-on-triangle-mesh
}.
\end{abstract}

\begin{keyword}
Triangle mesh \sep Hole detection \sep Boundaries formation \sep Underwater robotic \sep Multibeam sonar

\end{keyword}

\end{frontmatter}

\section{Introduction}
\label{sec:intro}
\begin{figure}[ht]
   \centering
    \begin{subfigure}[t]{0.48\textwidth}
        \centering
        \includegraphics[width=\textwidth]{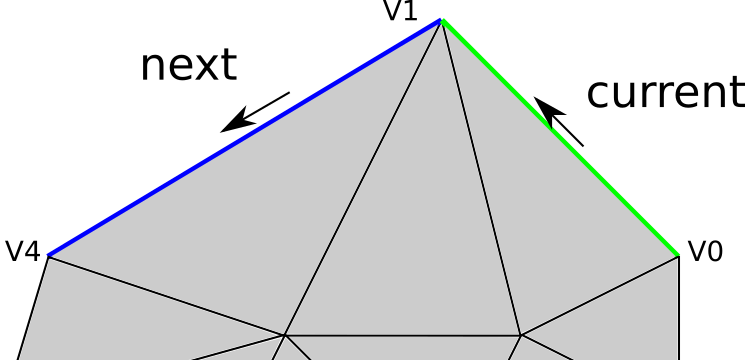}
        \caption{}
        \label{fig:problem_no}
    \end{subfigure}%
    ~ 
    \begin{subfigure}[t]{0.48\textwidth}
        \centering
        \includegraphics[width=\textwidth]{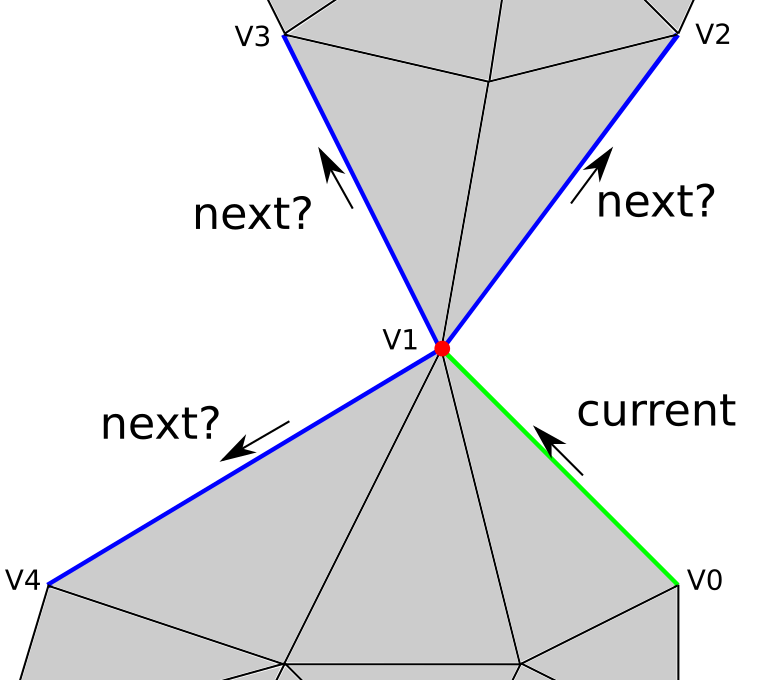}
        \caption{}
        \label{fig:problem_yes}
    \end{subfigure}%
    \caption{The green line represents the currently considered half-edge, while the blue lines indicate the potential next half-edges. The red dot in (b) indicates a singular vertex. In (a), determining the next half-edge is straightforward as only one blue line connects to the end of the current half-edge. However, in (b) a more complex scenario is presented with three potential next half-edges. The challenge lies in selecting the appropriate next half-edge to ensure that each half-edge in the entire mesh is used exactly once to construct boundaries, while also making sure that all half-edges contribute to boundary formation.}
    \label{fig:problem}
\end{figure}

In CAD, hole-detection methods are typically used as a preliminary step for hole-filling but often receive only peripheral attention. In fact, some studies about hole-filling even operate under the assumption that the holes are pre-identified or manually selected. However, automatic hole detection is crucial for several different application scenarios. Beyond its usefulness for CAD applications, it is of use for scene reconstruction from 3D point data acquired by robotic systems. Our use-case comes from the scene acquisition by autonomous underwater vehicles where the holes in the triangular meshes can be interpreted as information gaps during exploration missions. 
Most hole-filling algorithms directly adopt the method from \citet{liepa2003filling}, which assumes that the triangle mesh does not contain singular vertices.
However, singular vertices quickly appear in non-water-tight surface reconstructions generated by, for example, the Ball Pivoting Algorithm (BPA) (\citet{bernardini1999ball}).
Consequently, hole-detection based on \citet{liepa2003filling} will not detect all holes in triangle meshes containing singular vertices. The main challenge lies in selecting the appropriate next half-edge when the traversing method comes to a singular vertex and then ensuring that each half-edge in the entire mesh is traversed exactly once to construct boundaries and ensure that all constructed boundaries do not have any repeated vertices. 
This challenge becomes evident in the presence of singular vertices in a mesh, as depicted in \cref{fig:problem}.
Recently, the study by \citet{gou2022limofilling} aimed to solve this challenge, but their method necessitates a projection from 3D to 2D and requires a preprocessing step to modify the mesh under certain configurations before applying their hole-detection technique.

In this paper, we introduce a hole-detection method that reliably handles meshes with singular vertices without requiring any projections or modifications to the mesh. Furthermore, our method ensures that each detected hole will not contain repeated vertices, providing clear and concise information crucial for various applications, including underwater robotics, making it a reliable tool for handling complex mesh structures with singular vertices.

\cref{sec:related} presents a brief overview of relevant literature in the field. \cref{sec:preli} is dedicated to introducing and defining the specific terms used to explain our method.
Section \ref{sec:meth} explains how all boundaries can be formed from half-edges, even when they contain singular vertices. It also details how to partition boundaries with repeated vertices into multiple simple boundaries. Following this, we proposed a scheme to classify all different types of boundaries and offer a precise definition of a hole.
In \cref{sec:experiment}, we present three illustrative use cases of our method: two simulated and one derived from real-world data. Notably, each mesh model in these experiments has several singular vertices, yet our method consistently detects holes, irrespective of their presence.
The first experiment involves the application of our method to a well-known triangle mesh, the Stanford bunny mesh.
The second experiment demonstrates the application of our hole detection method on a simple triangular mesh from an underwater photogrammetry model.
The third experiment delves into a complex triangle mesh generated from a real-world point cloud produced by a multibeam sonar.
Our contribution is listed in \cref{sec:concl}.
\section{Related Work}\label{sec:related}
\citet{liepa2003filling} presents an elementary hole-detection algorithm assuming that the input triangular mesh is manifold, meaning it does not contain singular vertices, which are single vertices connected to more or less than two half-edges. This technique starts from a seed boundary vertex, tracing half-edges to identify closed loops. Since each boundary vertex is connected exactly by two half-edges, the tracing procedure is intuitive.

Other works considered with hole-filling, such as \citet{hu2012filling, jun2005piecewise, zhao2007robust}, and  \citet{qiang2010hole} adopt this hole-detection method, inherently carrying the same assumptions about the triangle mesh as presented in \cite{liepa2003filling}. Some hole-filling methodologies, such as those in \citet{li2010polynomial, wang2012hole, wu2022patch, hai2022cae}, do not explicitly provide details about their hole-detection approaches. This omission suggests a presumption that hole locations are already known, reflecting their primary aim to perform hole-filling rather than hole-detection.

A more recent paper by \citet{gou2022limofilling} describes a methodology that does not operate under the assumption that the mesh is perfectly manifold. Their described method introduces auxiliary segments as three-dimensional vectors, which are subsequently projected onto a 2D plane to check for overlaps with triangles. This solution can lead to modifications in the original triangle mesh and also depends on the chosen viewpoint for the projection from 3D to 2D, which is determined by the neighboring triangles of the half-edge associated with a singular vertex.

The study by \citet{feng2020fast} focuses on hole-filling for manifold meshes, implicitly assuming the meshes lack singular vertices. While they rely on the half-edge structure for hole detection, diverging from the approach in \cite{liepa2003filling}, which dominates most hole-filling research, they do not address the issue of singular vertices. The unclear connection between the number of 1-ring triangles and boundary detection made the replication of the described hole-detection technique infeasible. The half-edge data structure they employed assumes oriented meshes, while our proposed methodology functions without requiring the triangle mesh to be oriented, providing a more versatile solution.

We note that hole-detection is not typically the primary focus of works concerned with hole-filling methods, and, for example, in \citet{li2010polynomial, wang2012hole, wu2022patch, hai2022cae}, the method to detect holes is not mentioned at all.

For the task of classifying the holes and the main boundaries (model) from the boundaries, even if it may not be applicable to all meshes of objects, for surfaces that are relatively flat, the main boundary may be defined as the boundary with the largest length, a deviation from the method presented by \citet{qiang2010hole} of using the largest number of vertices. We offer a more fitting classification of the main (model) boundary and holes, drawing parallels to geographical terms such as tide-pool holes and lake holes.
\section{Preliminaries}
\label{sec:preli}
In the following, we define the technical terms used in this manuscript.
We first define the basic primitives like vertex, edge, and triangle. Then,
we define specific types of primitives.

\begin{definition}[\textbf{Vertex}]
A vertex $v$ is a single point located in 3D. Vertex $i$ is denoted as $v_i$.
\end{definition}

\begin{definition}[\textbf{Edge}]
An edge $e$ is a line segment that connects two different vertices. For orientation specification, when an edge connects from vertex $v_i$ to vertex $v_j$, the edge is denoted as $e_{ij}$
\end{definition}

\begin{definition}[\textbf{Triangle}]
A triangle $t$ is formed by interconnecting three vertices. Triangle $t_{ijk}$ is formed by the vertices $v_i$, $v_j$, and $v_k$.
\end{definition}

\begin{definition}[\textbf{Half-edge}] \label{def:half-edge}
A half-edge $h$ is an edge adjacent precisely to one triangle. For orientation specifications, when a half-edge connects from vertex $v_i$ to vertex $v_j$, the half-edge is denoted as $h_{ij}$.
\end{definition}

\begin{definition}[\textbf{Full-edge}]
A full-edge is an edge adjacent precisely to two triangles.
\end{definition}

\begin{definition}[\textbf{Mesh}]
A triangular mesh comprises a set $\BT$ of triangles that may be connected by their common edges or vertices.
\end{definition}

\begin{figure}[htb]
   \centering
    \begin{subfigure}[t]{0.32\textwidth}
        \centering
        \includegraphics[width=\textwidth]{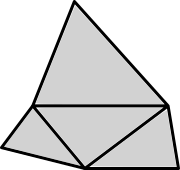}
        \caption{A edge-connected mesh.}
        \label{fig:mesh-single}
    \end{subfigure}%
    ~ 
    \begin{subfigure}[t]{0.32\textwidth}
        \centering
        \includegraphics[width=\textwidth]{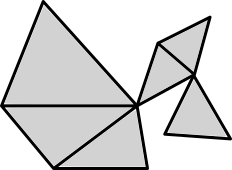}
        \caption{A vertex-connected mesh.}
        \label{fig:mesh-vertex}
    \end{subfigure}%
    \caption{Illustrating some of the mentioned mesh definitions. (a) Edge-connected mesh. (b) Three edge-connected meshes are connected by vertices to form a vertex-connected mesh.}
    \label{fig:mesh}
\end{figure}

\begin{definition}[\textbf{Edge-connected Mesh}]\label{def:edge-connected-mesh}
An edge-connected mesh consists of a set of triangles in which any two triangles connected by a vertex $v_i$ are also connected by another vertex $v_j$, see \cref{fig:mesh-single}.
\end{definition}

\begin{definition}[\textbf{Vertex-connected Mesh}]
A vertex-connected mesh is a set of triangles where at least two triangles are connected to each other only by a single vertex and do not share any common edges, see \cref{fig:mesh-vertex}.
\end{definition}

\begin{definition}[\textbf{Edge-manifold Mesh}]\label{def:edge-mesh}
An edge-manifold mesh is a triangle mesh with every edge adjacent to a maximum of two triangles.
\end{definition}

\begin{definition}[\textbf{Manifold Mesh}]
A manifold mesh is a triangle mesh that is both edge-manifold and vertex-manifold, meaning it contains no singular vertices. 
\end{definition}

\begin{definition}[\textbf{Boundary}]
A boundary is formed by half-edges connected consecutively to create a closed loop, denoted as $\Bb$.
\end{definition}

\begin{definition}[\textbf{Singular Vertex}]\label{def:singular}
A singular vertex is defined as a vertex to which more than two half-edges are connected.
\end{definition}

\begin{figure}[ht]
   \centering
    \begin{subfigure}[t]{0.45\textwidth}
        \centering
        \includegraphics[width=\textwidth]{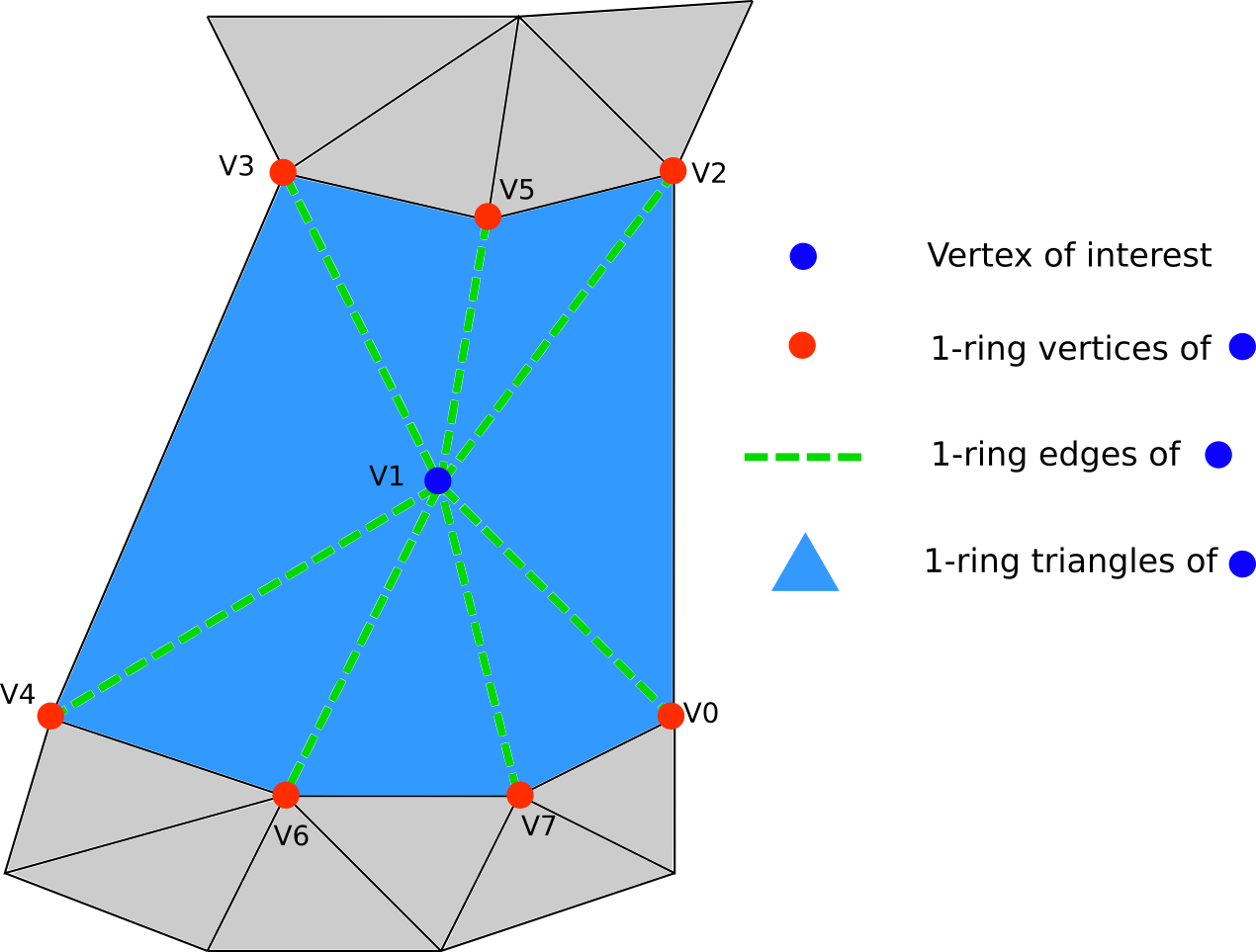}
        \caption{}
        \label{fig:basic:ring}
    \end{subfigure}%
    \quad \quad
    \begin{subfigure}[t]{0.35\textwidth}
        \centering
        \includegraphics[width=\textwidth]{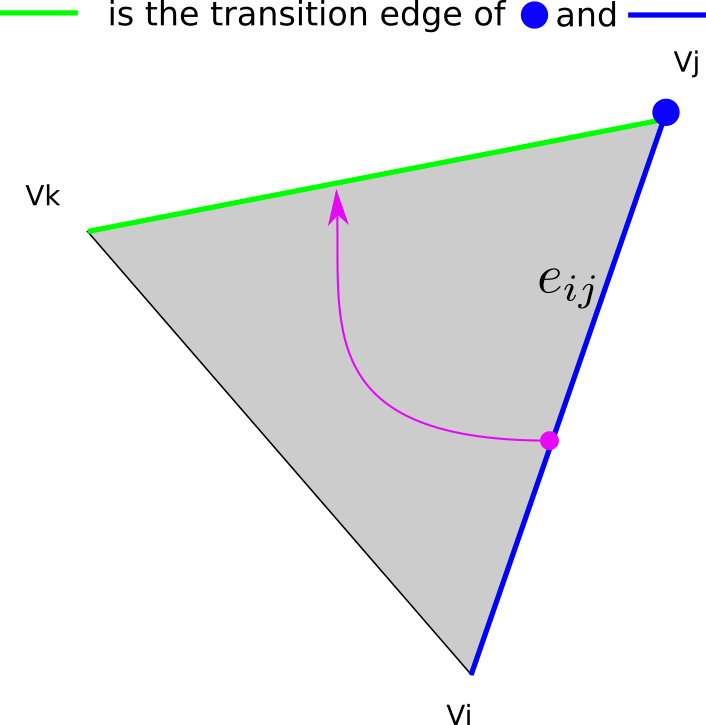}
        \caption{}
        \label{fig:basic:transition}
    \end{subfigure}%
    \caption{Illustrating the provided definitions of 1-ring triangles within (a) and a transition edge in (b).}
    \label{fig:basic}
\end{figure}

\begin{definition}[\textbf{1-ring Triangles}]\label{def:one-ring-triangles}
Given a vertex $v$, 1-ring triangles of $v$ build a set of triangles that are connected to vertex $v$, see \cref{fig:basic:ring}
\end{definition}

\begin{definition}[\textbf{Transition Edge}]\label{def:transition-edge}
Given a triangle $t_{ijk}$ and an oriented edge $e_{ij}$ (from vertex $v_i$ to $v_j$), the transition edge of $e_{ij}$ is an edge that has vertex $v_j$ but not $v_i$ connected, and it is one of the edges in $t_{ijk}$. This concept is depicted in \cref{fig:basic:transition}.
\end{definition}

We use curly brackets $\{...\}$ to represent a \textit{set}, following the convention from set theory; Square brackets $[...]$ denote an \textit{ordered array}, which maintains order and permits repeated elements; Angle brackets $\langle ... \rangle$ represents a \textit{cyclic array}.
\section{Methods and Algorithmic details}
\label{sec:meth}
The method presented in this paper originated from the practical problem of reconstructing 3D underwater scenes that are observed by an autonomous underwater vehicle (AUV). 
The fundamental concept is to interpret the gaps in the reconstructed mesh surface as topological gaps, representing areas where data are incomplete. Depending on their priority, these gaps require revisits by the AUV to gather additional data and complete the information in those regions. However, existing algorithms for hole detection in 3D meshes have demonstrated inadequacy in accurately identifying all gaps within obtained real-world 3D meshes, which have many singular vertices. Consequently, we developed a rigorous and theoretically well-founded algorithm to systematically determine all gaps in a 3D triangle mesh.

Given an edge-manifold triangle mesh (as defined in \cref{def:edge-mesh}), along with a collection of half-edges (as defined in \cref{def:half-edge}) within the mesh, our objectives are:
\begin{objective}\label{obj:1}
    To create boundaries from the half-edges in a manner that every half-edge will be exclusively used to construct one and only one boundary (ensuring existence and uniqueness). This does not mean that only one boundary will be constructed in a triangle mesh.
\end{objective}
\begin{objective}\label{obj:2}
    To ensure that there are no repeated vertices within each boundary. This is crucial for maintaining simplicity, a necessary feature when utilizing these boundaries as information gaps in underwater robotics. 
\end{objective}
\begin{objective}\label{obj:3}
    To classify main boundaries and holes from all detected boundaries.
\end{objective}

The complete set of half-edges can be easily obtained by a simple search or by more efficient algorithms readily available in computer graphics \cite{zhou2018open3d}. 

First, we explain how all boundaries, independent of the presence of singular vertices, are determined from the half-edge set. 
Then, we divide the boundaries that contain a repeated vertex into separate boundaries with no repeated vertices.
In addition, we distinguish main boundaries and holes and categorize them accordingly.
Finally, we explain how to analyze the characteristics of a hole to determine its location and dominant orientation. In our use case, a robot can utilize this knowledge to proactively ``fill up'' determined information holes by acquiring additional point clouds.

\subsection{Boundary (hole) detection}
\label{subsec:detection}
Given any edge-manifold mesh $\set T$, the set $\set H$ of half-edges can be acquired by searching through all triangles in $\set T$. Our process begins with $\set H$ and $\set T$.

\subsubsection{Stage 1/2: Finding boundaries regardless of singular vertex present.}
\label{subsubsec:connect_edges}
\begin{figure}[ht]
    \begin{subfigure}[t]{0.48\textwidth}
        \centering
        \includegraphics[width=\textwidth]{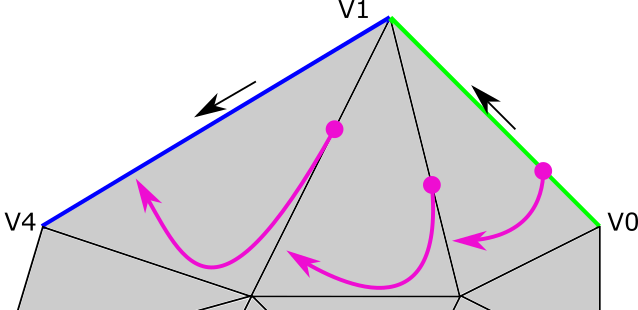}
        \caption{$v_1$ is non a singular vertex.}
        \label{fig:forming_boundary:b}
    \end{subfigure}%
    ~ 
   \centering
    \begin{subfigure}[t]{0.48\textwidth}
        \centering
        \includegraphics[width=\textwidth]{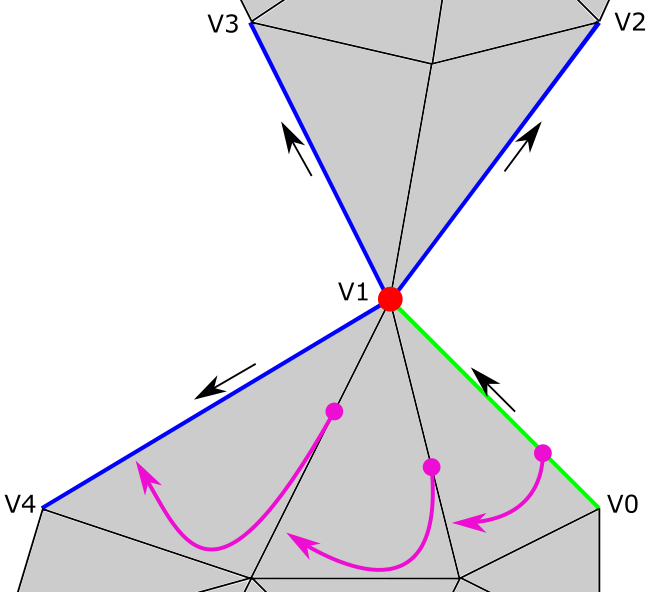}
        \caption{$v_1$ is a singular vertex.}
        \label{fig:forming_boundary:a}
    \end{subfigure}%
    \caption{A distinctive approach to determining the next connected half-edge to $h_{0,1}$ (green) is illustrated, irrespective of whether $v_1$ is singular or not. The purple curved arrow indicates the connected subsequent half-edge is determined.}
    \label{fig:forming_boundary}
\end{figure}

Our first objective is to establish a method for traversing all boundaries of the half-edge set $\set H$, even in the presence of singular vertices. This method ensures that each half-edge in the set ($h\in \set H$) corresponds to one and only one boundary $\set B$. The most significant challenge lies in determining the subsequently connected half-edge when the current half-edge involves a singular vertex.

For example, as shown in \cref{fig:forming_boundary:a}, let us consider the current half-edge $h_{0,1}$. There are three other half-edges that share vertex $v_1$, but not vertex $v_0$, and these are highlighted in blue.
The question that arises is: Which one of these should be selected as the next connected half-edge? This choice is crucial to ensure that each half-edge of the set $\set H$ is used exactly once to construct the boundaries.

To address this challenge, we need to establish a method to consistently determine the next connected half-edge when a singular vertex is involved. The purple arrows in \cref{fig:forming_boundary:a} illustrate the concept behind our solution to this issue. The core concept revolves around defining a unique orientation, represented by the purple arrows in \cref{fig:forming_boundary}, and iteratively following these purple arrows until a valid half-edge is encountered. \Cref{alg:next-halfedge} outlines how we identify the next half-edge. 
\begin{algorithm}[ht]
\caption{$h_{next} \gets next\_halfedge(h_{current},\set H, \set T).$}
\label{alg:next-halfedge}
$t \gets Find\_the\_triangle\_has\_half\_edge(h_{current}, \set T)$\;
$e_{current}\gets Find\_transition\_edge(h_{current}, t)$\;
\eIf{$e_{current} \in \set H$}
{
    $h_{next} \gets e_{current}$\;
    \Return $h_{next}$
}
{ 
    $e_{previous} \gets h_{current}$\;
    \While{\True}
    { 
        $t \gets Find\_t\_has\_first\_but\_not\_second\_edge(e_{current}, e_{previous}, \set T)$\; 
        $e_{next} \gets Find\_transition\_edge(e_{current}, t)$\;
        \eIf{$e_{next} \in \mathbf{h}$}
        {
            $h_{next} \gets e_{next}$\;
            \Return $h_{next}$
        }
        {
            $e_{previous} \gets e_{current}$\;
            $e_{current} \gets e_{next}$\;
        }
    }
}
\end{algorithm}

\begin{figure}[ht]
   \centering
    \begin{subfigure}[t]{0.45\textwidth}
        \centering
        \includegraphics[width=\textwidth]{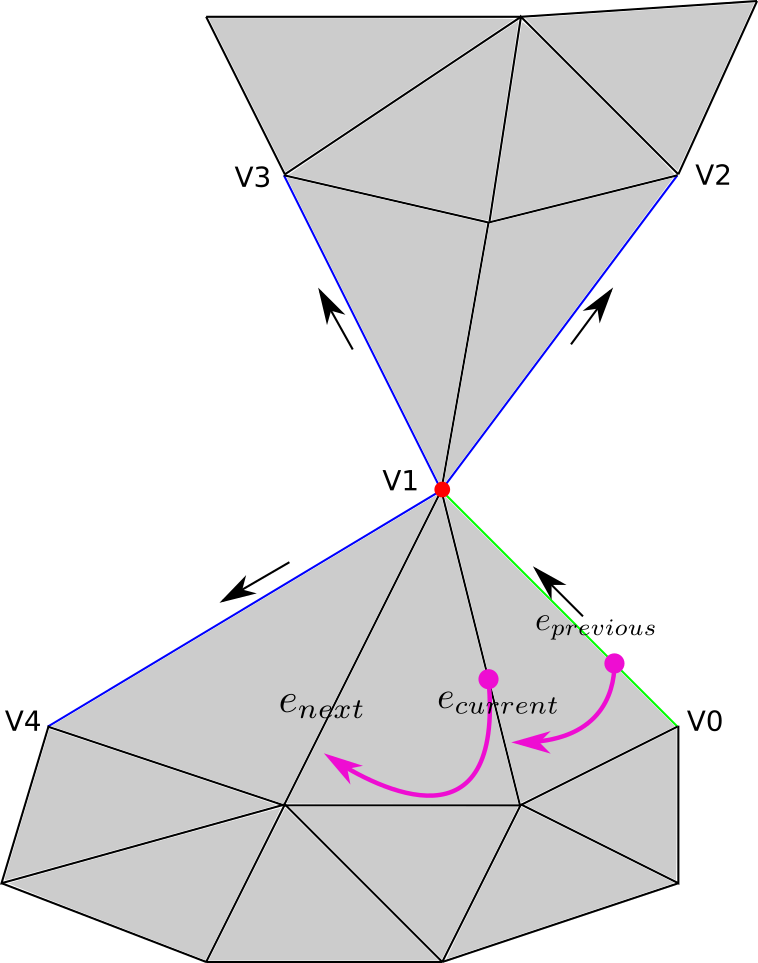}
        \caption{}
    \end{subfigure}%
    ~ 
    \begin{subfigure}[t]{0.45\textwidth}
        \centering
        \includegraphics[width=\textwidth]{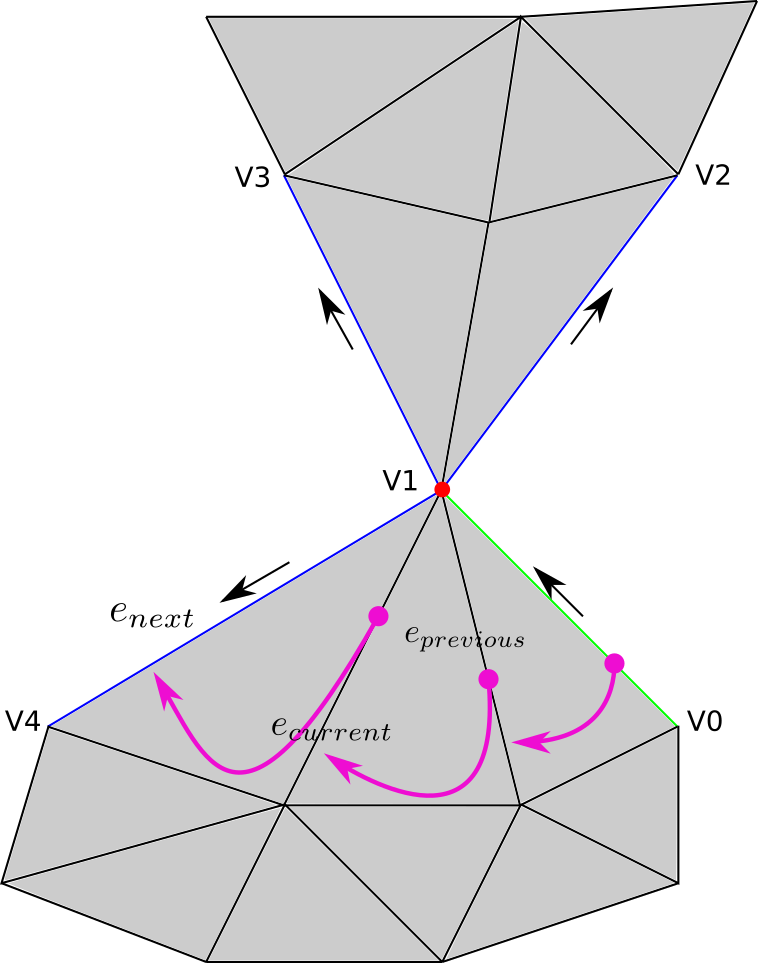}
        \caption{}
    \end{subfigure}%
    \caption{The looping process (line 8 - 17) in \Cref{alg:next-halfedge}.}
    \label{fig:next-edge_looping}
\end{figure}

The following is a step-by-step explanation of \Cref{alg:next-halfedge}.
\begin{itemize}
    \item \Cref{alg:next-halfedge} requires as input the current half-edge $h_{current}$, the half-edge set ($\set H$), and the triangle set $\set T$.
    \item Line 1: Find the triangle $t$ containing the half-edge $h_{current}$. According to the definition of a half-edge, precisely one triangle $t \in \set T$ is associated with the half-edge $h_{current}$.
    \item Line 2: Determine the transition edge $e_{current}$ of $h_{current}$ and $t$. The orientation of $h_{current}$ will be utilized (see \cref{def:transition-edge}).
    \item Lines 3 - 5: If $e_{current}$ is a half-edge, we have successfully identified the half-edge subsequently connected to the current half-edge $h_{current}$. Return $e_{current}$ as $h_{next}$ and terminate the algorithm.
    \item Lines 7 - 18: If $e_{current}$ is not a half-edge, then $e_{previous}$ is assigned as $h_{current}$, and the algorithm enters a loop spanning lines 8 to 18. It is important to note that $e_{current}$ must be a full-edge, given that it is not a half-edge and $\set T$ adheres to the edge-manifold assumption.
    \item Line 9: Find the triangle $t$ containing the full-edge $e_{current}$, while excluding the presence of the edge $e_{previous}$.
    Since $\set T$ is an edge-manifold mesh and $e_{current}$ is a full-edge, there exists precisely one triangle that contains the full-edge $e_{current}$, but not the edge $e_{previous}$. See \cref{fig:next-edge_looping}.
    \item Line 10: Determine the transition edge $e_{next}$ of $c_{current}$ and $t$, as illustrated in \cref{fig:next-edge_looping}. 
    \item Line 11 - 13: If $e_{next}$ is a half-edge, we have successfully identified the subsequently connected half-edge to the current half-edge $h_{current}$. Return $e_{next}$ as $h_{next}$ and terminate the algorithm.
    \item Lines 15 - 16: In the event that $e_{next}$ is not a half-edge, proceed by reassigning $e_{previous}$ to $e_{current}$, and then reassign $e_{current}$ to $e_{next}$. Subsequently, return to line 9 to continue the process.
\end{itemize}
We have established a method to identify the next connected half-edge, based on the current half-edge, as outlined in \Cref{alg:next-halfedge}. Now, we explain the process of constructing boundaries from the set of half-edges $\set H$. The corresponding pseudocode is presented in \Cref{alg:boundaries-halfedges}, and the explanation of this pseudocode is provided in the following:
\begin{algorithm}[ht]
\caption{$\mathbf B \gets construct\_boundaries(\set T, \set H)$ }
\label{alg:boundaries-halfedges}
/* Create empty set $\set B$ */\;
$\set B \gets \{\,\}$\;
\While{$\set H$ is not empty}
{
    $h_{start} \gets \set H.random\_select()$ /* Random copy a element from the set $\set H$ */\;
    $\mathbf b \gets [ \,]$                        /* Create an empty ordered array\; 
    $\mathbf b.append(h_{start})$\;
    $h_{current} \gets h_{start}$\;
    \While{\True}
    {
        $h_{next} \gets next\_half\_edge(h_{current}, \set H, \set T)$\;
        $h_{next} \gets re\-orientation(h_{current}, h_{next})$\;
            \eIf{$h_{next}$ and $h_{start}$ are the same edge}
            {\Break}
            {
            $\mathbf b.append(h_{next})$\;
            $h_{current} \gets h_{next}$
            }
    }
    $\set B.insert(\mathbf b)$\;
    $\set H.remove(\mathbf b)$
}
\Return $\mathbf B$
\end{algorithm}
\begin{itemize}
    \item The input for \Cref{alg:boundaries-halfedges} consists of the set of half-edges $\set H$ and an edge-manifold triangle mesh $\set T$.
    \item Line 2: Initialize an empty set $\set B$ that will be used to store the boundaries.
    \item Line 4: Randomly choose a half-edge from $\set H$ and assign it as $h_{start}$. This selected half-edge $h_{start}$ will serve as the termination criterion, indicating the completion of a closed loop of connected half-edges constituting a boundary. Note that $h_{start}$ is not removed from $\set H$ at this stage.
    \item Line 5: Generate an empty ordered array ($\set b$), which will be utilized to accumulate connected half-edges and ultimately assemble a boundary.
    \item Line 6: Assign the current half-edge $h_{current}$ as $h_{start}$.
    \item Line 9: Determine the next connected half-edge using the procedure described in \Cref{alg:next-halfedge}.
    \item Line 10: Re-orient $h_{next}$ such that its starting vertex is equal to the ending vertex of $h_{current}$.
    \item Line 11: If $h_{next}$ is identical to $h_{start}$, a boundary has been formed, leading to the termination of the while-loop at line 8.
    \item Line 14: Add $h_{next}$ to the ordered array $\set b$.
    \item Line 15: Re-assign $h_{current}$ as $h_{next}$.
    \item Line 18: A boundary is discovered, and insert the boundary $\set b$ into $\set B$. 
    \item Line 19: Remove all half-edges that belong to the boundary $\set b$ of the set of half-edges $\set H$.
    \item Line 20: If $\set H$ is not empty, back to Line 4.
    \item Line 21: The half-edge set $\set H$ is now empty and the boundary set $\set B$ is returned.
\end{itemize}
Combining \Cref{alg:boundaries-halfedges} and \Cref{alg:next-halfedge}, each half-edge $h$ present in the half-edge set $\set H$ is used to create a single boundary, ensuring its existence and uniqueness. This statement is proven in \cref{thm:next-edge} in \ref{app:thm}, which is further substantiated by a corresponding mathematical proof in \ref{app:thm}. 

Diverging from the approach presented in \citet{gou2022limofilling}, our methodology possesses the ability to handle singular vertices without necessitating any form of 3D to 2D projection and no auxiliary segments as introduced in \citep{gou2022limofilling}. 
Additionally, we avoid the need for any pre-processing steps to alter the mesh. 
Contrary to the standard boundary (holes) detection method in \cite{liepa2003filling}, which assumes that there are no singular vertices in the triangle mesh, our method can identify boundaries even in the presence of singular vertices.
Our solution to this challenge is based on the inclusion of full-edge information from $\set T$ (cf. \Cref{alg:next-halfedge} line 9). As a result, the presence of singular vertices along the half-edge $\set T$ becomes irrelevant. \Cref{alg:boundaries-halfedges} successfully achieves our \cref{obj:1}.

\subsubsection{Stage 2/2: Decomposition of a complex boundary into simple boundaries}
\label{subsubsec:decompose}
\begin{figure}
   \centering
    \begin{subfigure}[t]{0.49\textwidth}
        \centering
        \includegraphics[width=\textwidth]{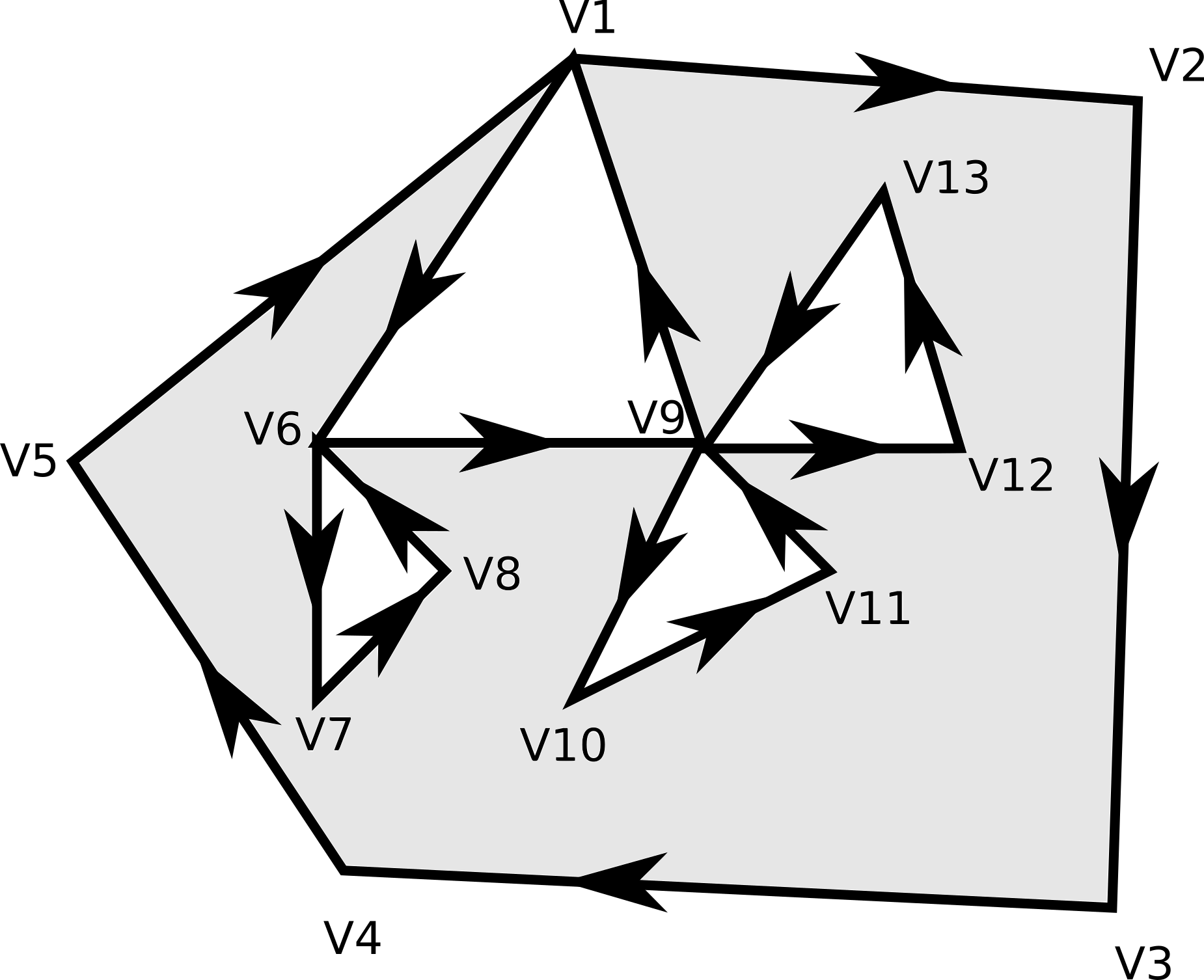}
        \caption{A single boundary with repeated vertices (a complex boundary).}
        \label{fig:boundary-manifoldness-a}
    \end{subfigure}%
    ~ 
    \begin{subfigure}[t]{0.49\textwidth}
        \centering
        \includegraphics[width=\textwidth]{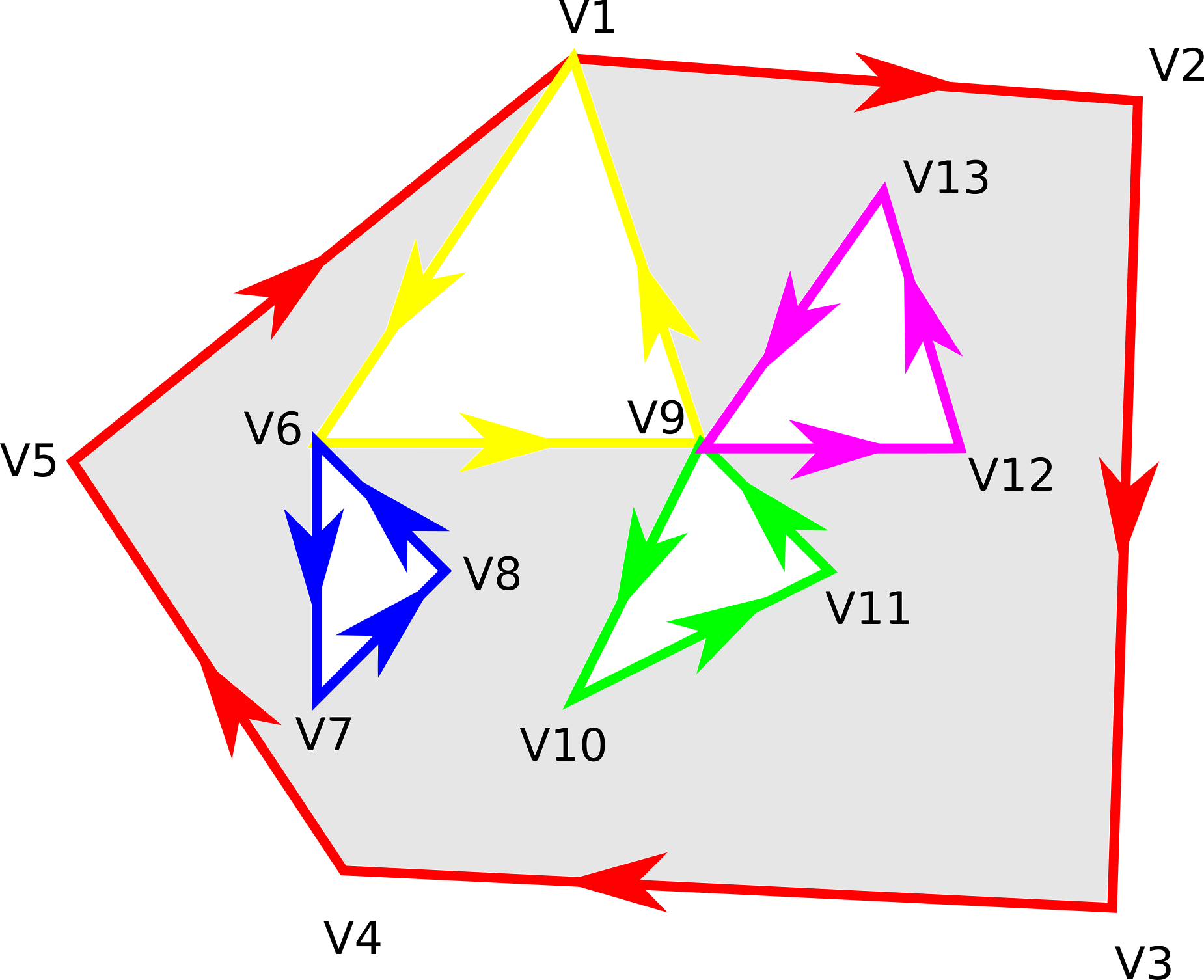}
        \caption{Five different simple boundaries without repeated vertices.}
        \label{fig:boundary-manifoldness-b}
    \end{subfigure}
    \caption{Complex boundary decomposition. The gray background indicates the rest of the triangular mesh.}
    \label{fig:boundary-manifoldness}
\end{figure}

As described in \cref{subsubsec:connect_edges}, we have derived a set of boundaries denoted as $\mathbf{B}$. It is important to note that the boundary attained through the procedure detailed in \Cref{alg:boundaries-halfedges} may contain duplicated vertices. In light of this, we establish the following definition:
\begin{definition}[\textbf{Simple boundary}]
    A simple boundary is a boundary that has no repeated vertices.
\end{definition}
\begin{definition}[\textbf{Complex boundary}] \label{def:com_bound}
    A complex boundary is a boundary that has repeated vertices.
\end{definition}
For example, in \cref{fig:boundary-manifoldness-a}, we have a complex boundary with vertices order: \[\langle v_5,v_1,v_6,v_7,v_8,v_6,v_9,v_{10},v_{11},v_9, v_{12}, v_{13}, v_9, v_1,v_2,v_3,v_4 \rangle.\]
Vertex $v_1$ and $v_6$ occur twice each, while vertex $v_9$ has occurred three times.
Our second aim is to decompose this complex boundary into several simple boundaries\footnote{In graph theory, a complex boundary is a \textbf{Euler circuit \cite[Definition 11.15, Definition 11.3]{grimaldi2006discrete}}, and we want to decompose the Euler circuit into \textbf{circle(s) \cite[Definition 11.3]{grimaldi2006discrete}}, which is a simple boundary.}, like \cref{fig:boundary-manifoldness-b};
In this case, our desired outcomes are $\langle v_5,v_1,v_2,v_3,v_4 \rangle$, $\langle v_1, v_6, v_9 \rangle$, $\langle v_{10}, v_{11}, v_9 \rangle$, $\langle v_6, v_7, v_8 \rangle$, $\langle v_9, v_{12}, v_{13} \rangle$, as indicated by the various colors in \cref{fig:boundary-manifoldness-b}.

The solution involves iteratively decomposing the complex boundary by dividing it into two separate boundaries whenever repeated vertices are encountered. This process is continued until no repeated vertices remain.
For instance, let's consider a boundary depicted as shown in \cref{fig:boundary-manifoldness-a}, which we can represent as a half-edge ordered array:
\begin{multline}    
[h_{5,1},h_{1,6},h_{6,7},h_{7,8},h_{8,6},h_{6,9},h_{9,10},h_{10,11},h_{11,9}, \\
h_{9,12}, h_{12,13},h_{13,9},h_{9,1},h_{1,2},h_{2,3},h_{3,4},h_{4,5}]
\end{multline}
In the process of partitioning the complex boundary, we select any instance of repeated vertices. In the current situation, we arbitrarily choose vertex $v_9$ for decomposition.
We separate the complex boundary $\set b$ into three segments, illustrated by the colors red, blue, and green; see \cref{eq:decom}.
The transition from red to blue occurs the first time two half-edges are linked via the vertex $v_9$, while the transition from blue to green takes place the second time two half-edges are connected through the vertex $v_9$.
The following expression shows the result:
\begin{multline}
    [
    {\color{red}h_{5,1},h_{1,6},h_{6,7},h_{7,8},h_{8,6},h_{6,9},}
    {\color{blue}h_{9,10},h_{10,11},h_{11,9}}, \\
    {\color{green}h_{9,12},h_{12,13},h_{13,9},h_{9,1},h_{1,2},h_{2,3},h_{3,4},h_{4,5}}
    ].
    \label{eq:decom}
\end{multline}
(Note that the color of the above expressions is not related to the colors used in \cref{fig:boundary-manifoldness-b}.)
We define $\set b_1$ as
\begin{equation}
[
{\color{red}h_{5,1},h_{1,6},h_{6,7},h_{7,8},h_{8,6},h_{6,9},}
{\color{green}h_{9,12},h_{12,13},h_{13,9},h_{9,1},h_{1,2},h_{2,3},h_{3,4},h_{4,5}}
]
\end{equation}
by combining the red and green array.
Additionally, we define $\set b_2$ as:
\begin{equation}
[
{\color{blue}h_{9,10},h_{10,11},h_{11,9}}
].
\end{equation}
$\set b_2$ is a simple boundary since it does not have a repeated index (cf. \cref{fig:boundary-manifoldness-b}). 
However, $\set b_1$ remains a complex boundary due to the recurrence of repeated indices $v_6$, $v_1$, and $v_9$. In particular, $v_9$ now recurs only twice instead of three times. 
To further decompose $\set b_1$, we repeat the same process recursively until we successfully break down all these boundaries into simpler boundaries, as visualized in \cref{fig:boundary-manifoldness-b}.
As stipulated by \cref{thm:decom} in \ref{app:thm}, the procedure guarantees that we can systematically decompose a complex boundary into two boundaries, each containing fewer half-edges than the original complex boundary. This ensures that we will ultimately derive multiple simple boundaries from a complex one.
The pseudo-code is shown in \Cref{alg:decomposition}. $\set S$ is a set that contains all the simple boundaries.

\begin{algorithm}[ht]
\caption{$\set S \gets decompose\_complex\_to\_simples(\set b)$}
\label{alg:decomposition}
\eIf{$has\_repeated\_vertex(\set b)$}
{
    $index\_1, index\_2 \gets Find\_repeated\_index(\set b)$\;
    $\set b_1 \gets \set b[:index\_1] + \set b[index\_2:]$\;
    $\set b_2 \gets \set b[index\_1:index\_2]$\;
    $\set S_1 \gets decompose\_complex\_to\_simples(\set b_1)$\;
    $\set S_2 \gets decompose\_complex\_to\_simples(\set b_2)$\;
    \Return $\set S_1 \cup \set S_2$
}
{
    $\set S \gets \{\set b\}$\;
    \Return $\set S$
}
\end{algorithm}
The explanation of \Cref{alg:decomposition} is as follows:
\begin{itemize}
    \item Line $1$: This step involves checking whether the boundary $\set b$ contains repeated vertices.
    \item Line $2$: Randomly selects a repeated vertex, denoted as $v_r$, which appears more than once within $\set b$. If multiple repeated vertices exist, one is arbitrarily chosen as $v_r$. The algorithm then identifies the indices where $v_r$ repeats itself within $\set b$. If $v_r$ is repeated more than twice, two indices are chosen, ensuring that $index_1$ is smaller than $index_2$.
    \item Line $3$: The ``+'' symbol signifies the concatenation of two arrays.
    \item Lines $5-6$: The function is called recursively, resulting in the sets $\set S_1$ and $\set S_2$, both containing several (at least one) simple boundaries.
    \item Line $7$: The union operation combines two sets, $\set S_1$ and $\set S_2$.
    \item Line $9$: The set $\set S$ contains only one simple boundary $\set b$.
\end{itemize}
\Cref{alg:decomposition} aims to iteratively break down a complex boundary into multiple simple boundaries by selecting and detaching repeated vertices, ultimately identifying multiple simple boundaries. \Cref{alg:decomposition} successfully achieves our \cref{obj:2}. Note that even though \Cref{alg:boundaries-halfedges} line 4 and \Cref{alg:decomposition} line 2 have randomness, the final result is not random at all.

\subsection{Categorize holes from boundaries}
\label{subsec:cat_holes}
\begin{figure}[ht]
    \centering
    \includegraphics[width=0.7\textwidth]{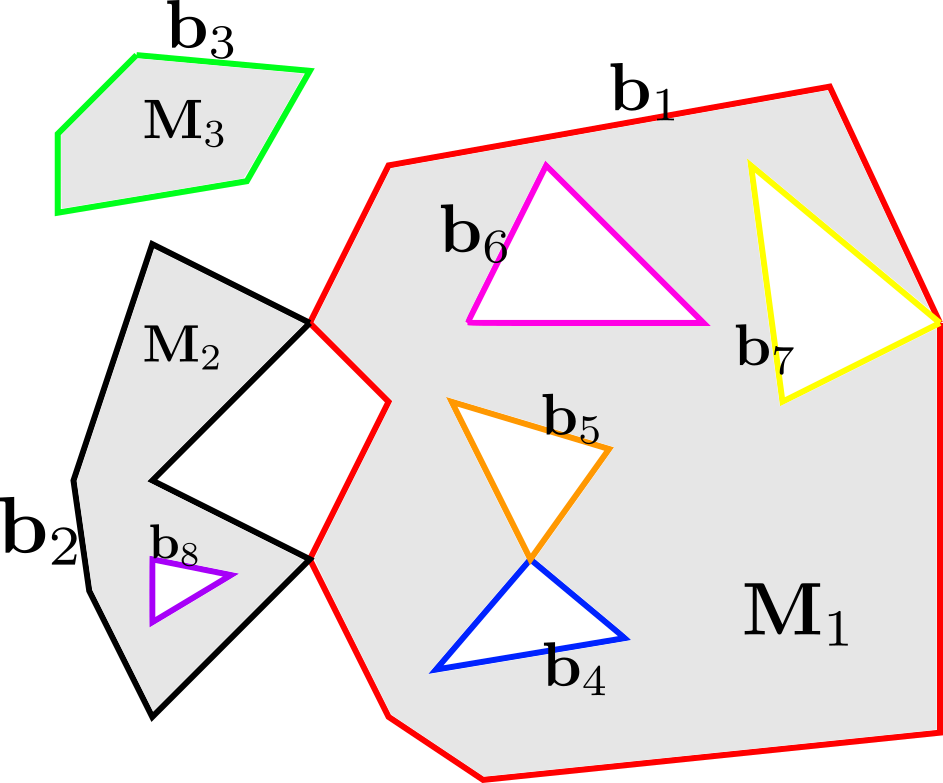}
    \caption{An example mesh with detected boundaries. The gray background indicates the rest of the triangular mesh.}
    \label{fig:mesh-bpa}
\end{figure}

\begin{figure}[ht]
    \centering
    \includegraphics[width=0.7\textwidth]{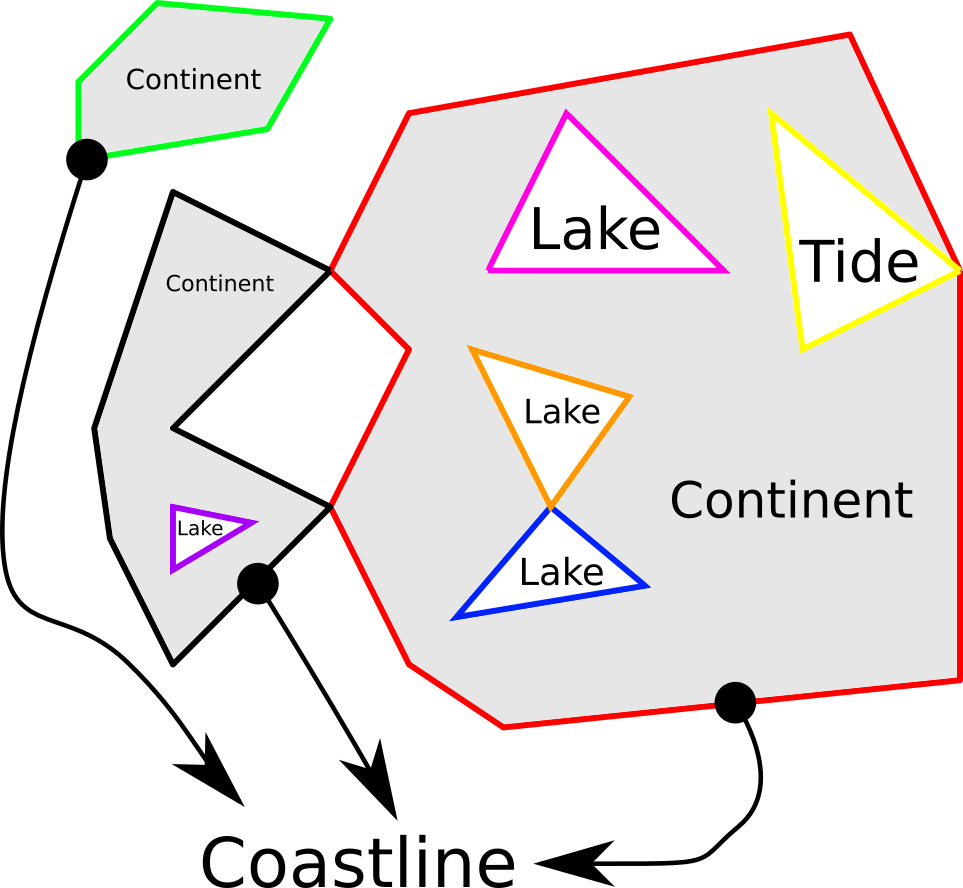}
    \caption{For mnemonic, we use geographic terms to name these types of boundaries. The red, black, and green boundaries can be seen as coastlines of three distinct continents, with each continent being an edge-connected mesh. The yellow boundary resembles a tidal hole as it connects with the red coastline. The pink, orange, and blue boundaries are considered to be lake holes within the red coastline since their edges belong to that same continent. Similarly, the purple hole is like a lake within the black coastline. The gray background indicates the rest of the triangular mesh.}
    \label{fig:land-ocean}
\end{figure}

Based on the information provided in \cref{subsec:detection}, all boundaries are constructed using the half-edge set $\set H$ and subsequently decomposed into simpler boundaries denoted as $\set S$.
To treat holes as regions with a lack of information, it is not suitable to classify all boundaries within $\set S$ as holes. 
Instead, some boundaries in $\set S$ are more suitable for being identified as the main boundaries, as demonstrated in \cref{fig:mesh-bpa}. On the contrary, certain boundaries in $\set S$ will indeed be considered as holes. We categorize the main boundaries and holes within $\set S$ as follows:
\begin{enumerate}
    \item If $\set S$ is not empty.
    \begin{enumerate}
        \item Extract and remove the boundary with the greatest length (sum of its edges) from $\set S$ and denote it as the coastline (main boundary) $\set c_i$.
        \item Determine the edge-connected mesh $\set M_i$ corresponding to $\set c_i$. We call this edge-connected mesh the continent of coastline $\set c_i$.
        \item Extract and remove all boundaries from $\set S$ if they share the same half-edges of the triangles present in $\set M_i$. These boundaries are considered holes.
        \item Check the holes. If a hole shares the same vertice with $\set c_i$, the hole will be categorized as a tide-pool hole; we will just call it a tide hole, denoted as $\set P_i$.
        \item The rest of the holes are classified as lake holes, indicated as $\set L_i$.
    \end{enumerate}
    \item Increment the index $i$ by 1.
    \item Repeat step 1 unless $\set S$ is empty.
\end{enumerate}
When applying the previously mentioned methodology to the scenario depicted in \cref{fig:mesh-bpa}, the outcomes are showcased in \cref{tab:categorization}. For mnemonic, we use geographic terms (i.e. coastline, tide-hole, lake) to name these types of boundaries; see \cref{fig:land-ocean}. We will continue to use the term coastline instead of the term main boundary.
The method in \cite{qiang2010hole} defines coastlines based on the number of vertices in a boundary. Using their definition, a boundary, even with a minimal length, can possess a large number of vertices. Therefore, we determine the coastlines using the maximum length of the boundaries, as we consider this to be more suitable.
As a result, the mesh $\set T$ in \cref{fig:mesh-bpa} is separated into three distinct edge-connected meshes $\set M_1$, $\set M_2$, and $\set M_3$, delineating three distinct continents. This segmentation is advantageous for underwater robotics since the robot can focus on the largest continent and on filling the lake hole(s) and tide hole(s) on the largest continent and initially ignore small ones. This classification procedure achieves our \cref{obj:3}.

\begin{table}[h]
\centering
\begin{tabular}{|c|c|c|c|}
\hline
 & Continent 1 & Continent 2 & Continent 3 \\ \hline
Coastline & $\set b_1$ & $\set b_2$ & $\set b_3$ \\ \hline
Edge-connected mesh & $\set M_1$ & $\set M_2$ & $\set M_3$ \\ \hline
Tide hole(s) & $\set P_1 = \{\set b_7 \}$ & $\set P_2 =\emptyset$ & $\set P_3 = \emptyset$ \\ \hline
Lake hole(s) & $\set L_1 = \{ \set b_4, \set b_5, \set b_6\}$ & $\set L_2 = \{\set b_8 \}$ & $\set L_3 = \emptyset$ \\ \hline
\end{tabular}
\caption{Categorization of holes in the example shown in \cref{fig:mesh-bpa}. The symbol $\emptyset$ represents an empty set.}
\label{tab:categorization}
\end{table}
\section{Implementation}
The implementation was carried out using Python \cite{python}.  Open3D's \cite{zhou2018open3d} python packages were used to load a triangle mesh in the ply file format and obtain the half-edge set $\BH$.
Our implementation also tests if the processed triangle is edge-manifold.
If the mesh is edge-manifold, it begins by extracting boundaries. These boundaries have been decomposed from complex forms, ensuring that all are presented as simple boundaries. A secondary output provides the relationship between coastlines (main boundaries), continents (edge-connected meshes), tide holes, and lake holes, as shown in \cref{tab:categorization}. Due to its computational intensity, the secondary output can be disabled, especially if users are primarily interested in the initial boundary information. All outputs are saved in a JavaScript Object Notation (JSON) file.

When vertex $v_j$ is not a singular vertex, our optimization involves searching for the next connected half-edge of $h_{i,j}$ directly within the $\BH$ structure, circumventing the need to search within the entirety of $\BT$. The source code can be found in the abstract. 

\section{Experimental results}
\label{sec:experiment}
We tested three distinct cases with our method. In the first, we applied our method to a well-known 3D triangle mesh, the Stanford bunny \cite{bunny} mesh obtained from Open3D. The second case examined our method on an underwater photogrammetry model, presenting a simple triangular mesh with singular vertices. Finally, in the third case, we utilized a real dataset, demonstrating our method's capability to detect holes on intricate surfaces derived from real point clouds.

\subsection{The bunny triangles mesh}
\begin{figure}[htbp]
    \centering
    \begin{subfigure}[t]{0.490\textwidth}
        \includegraphics[width=\textwidth]{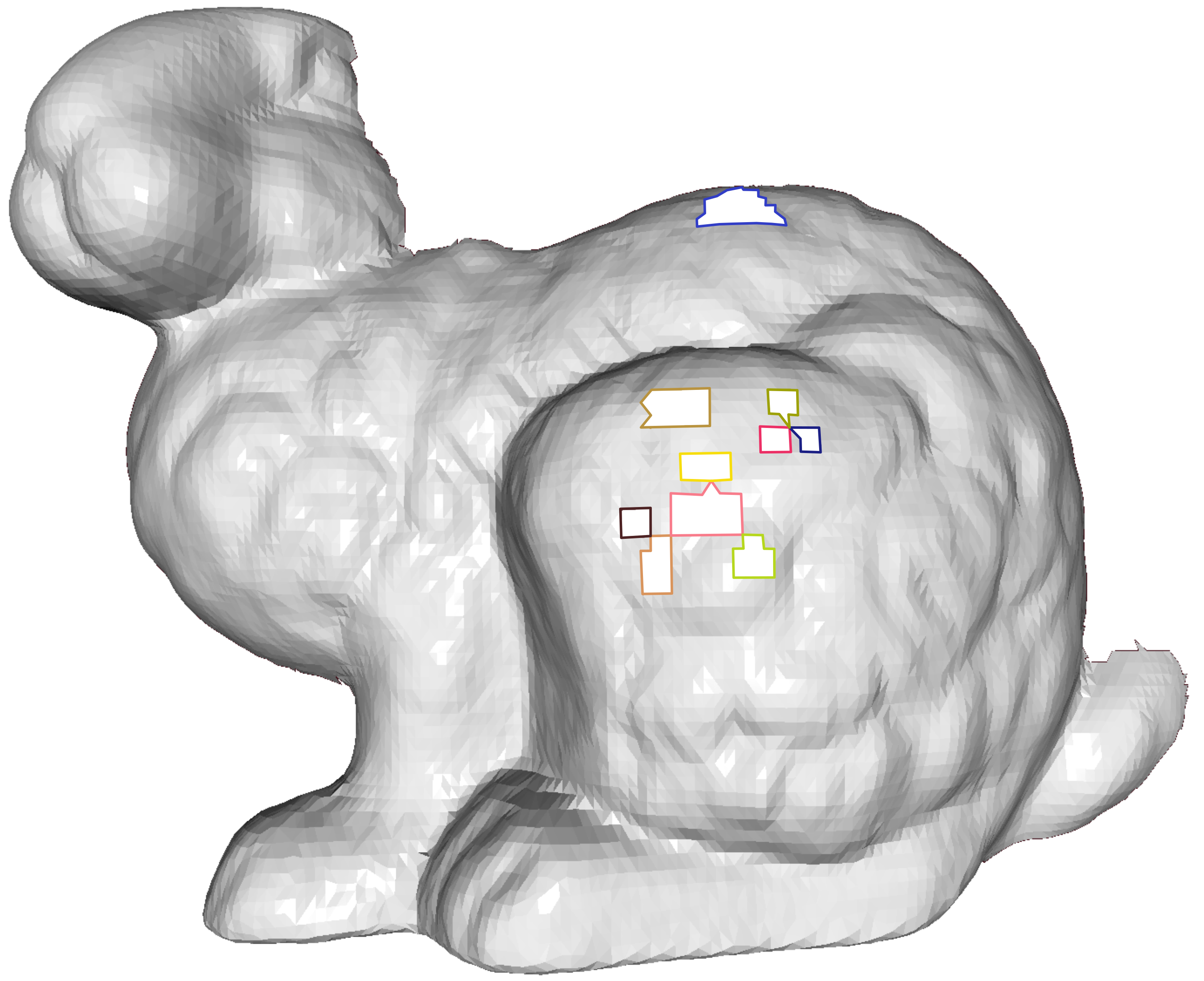}
        \caption{Detected holes}
    \end{subfigure}
    \begin{subfigure}[t]{0.490\textwidth}
        \includegraphics[width=\textwidth]{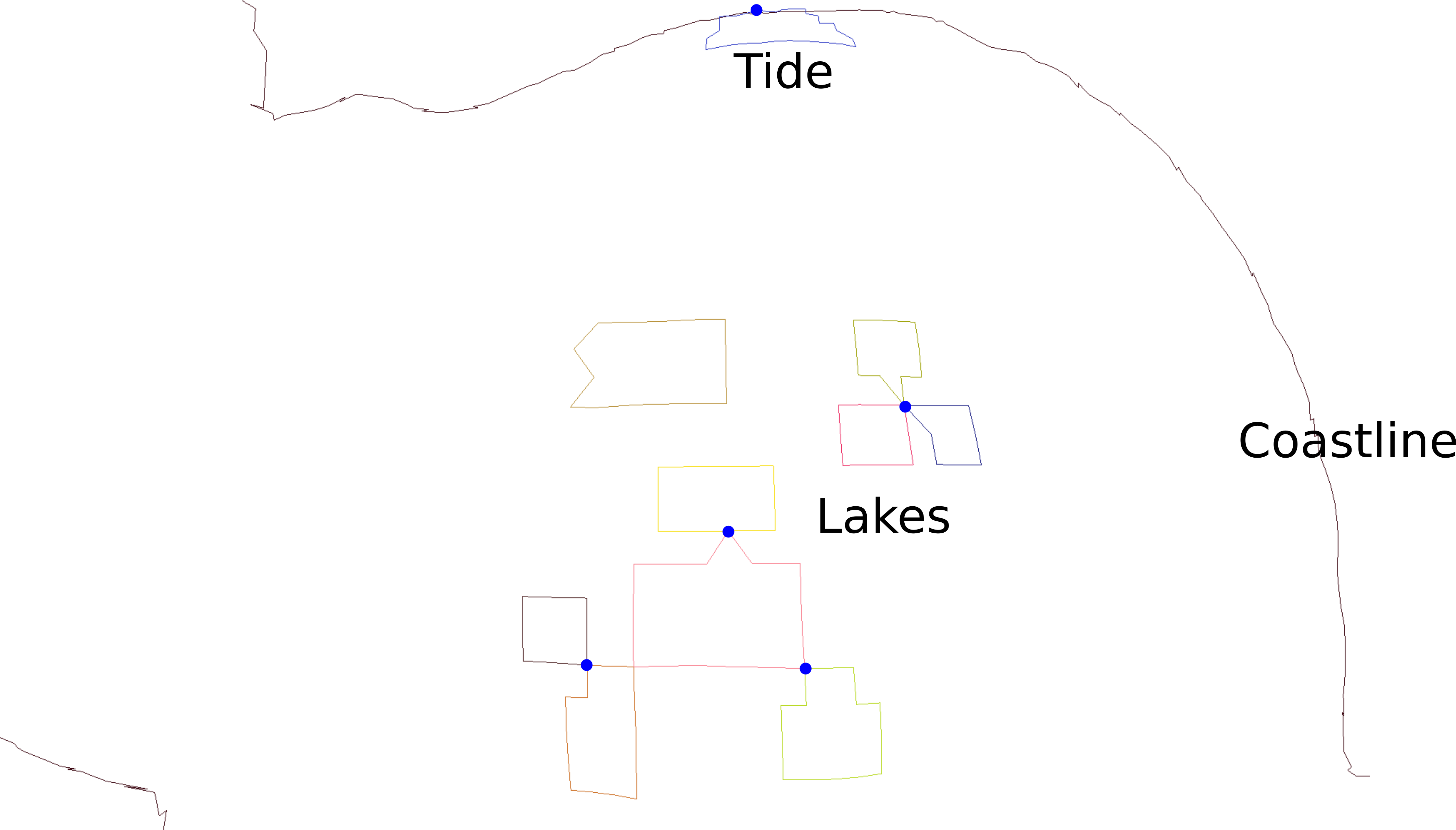}
        \caption{One coastline; One tide hole; The rest are lake holes; Blue dots are singular vertices.} 
    \end{subfigure}
    \caption{Testing our hole detection method on the bunny mesh.}
    \label{fig:bunny_holes}
\end{figure}

We evaluated our method on the Stanford bunny \cite{bunny} triangle mesh. To improve the clarity of boundary visualization, the original mesh was split in half, as illustrated in \cref{fig:bunny_holes}.
Holes were manually introduced, resulting in a mesh model that contained four singular vertices. Despite the presence of singular vertices, our method successfully detected all boundaries, identifying one coastline (main boundary), one tide hole, and nine lake holes. A closer examination of \cref{fig:bunny_holes} reveals that only one boundary has no singular vertices. This indicates that using the conventional hole detection technique by \citep{liepa2003filling}, only one boundary would have been detected.

\begin{figure}[htbp]
    \centering
    \begin{subfigure}[t]{0.490\textwidth}
        \includegraphics[width=\textwidth]{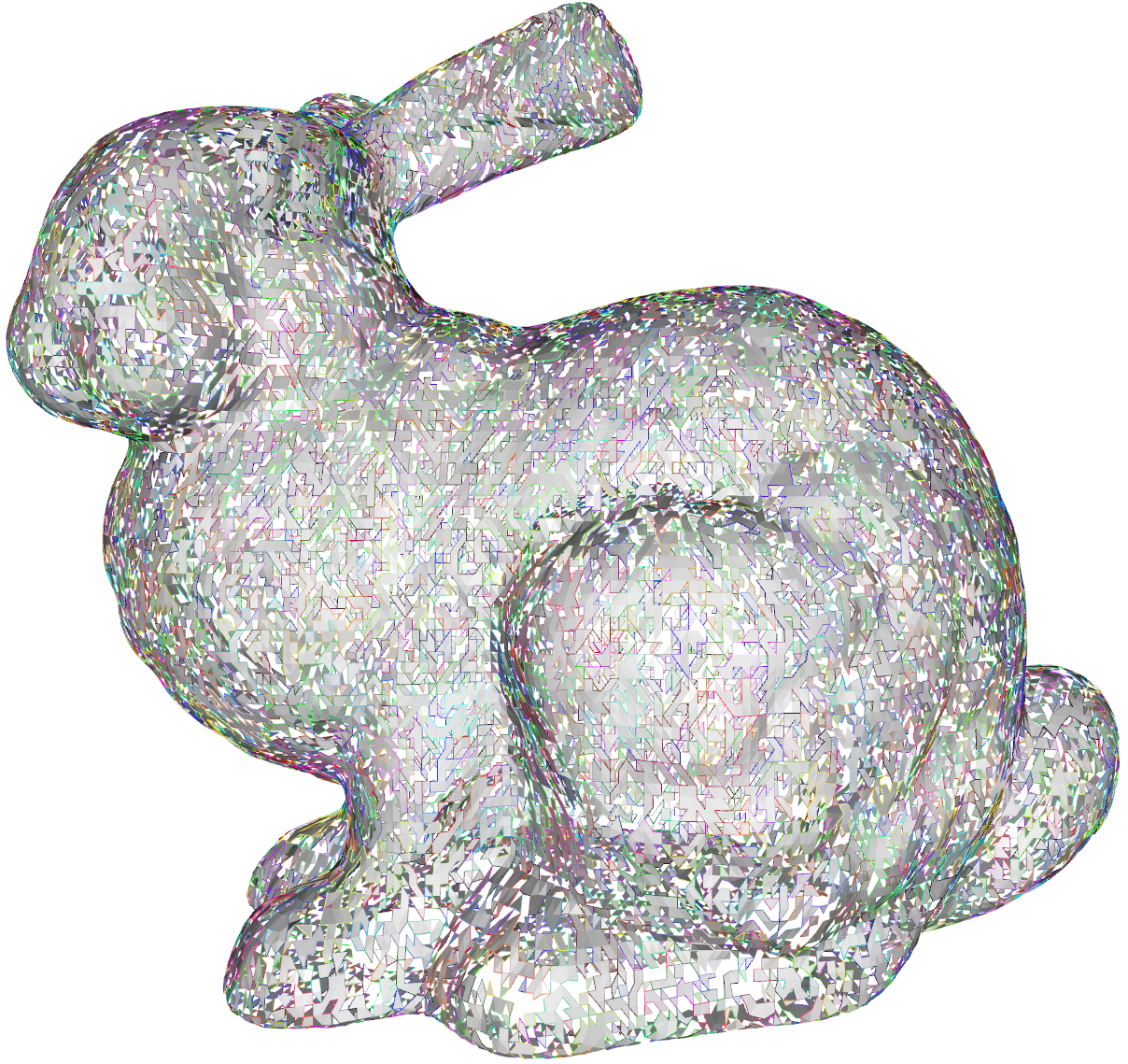}
        \caption{Boundaries with mesh.}
    \end{subfigure}
    \begin{subfigure}[t]{0.490\textwidth}
        \includegraphics[width=\textwidth]{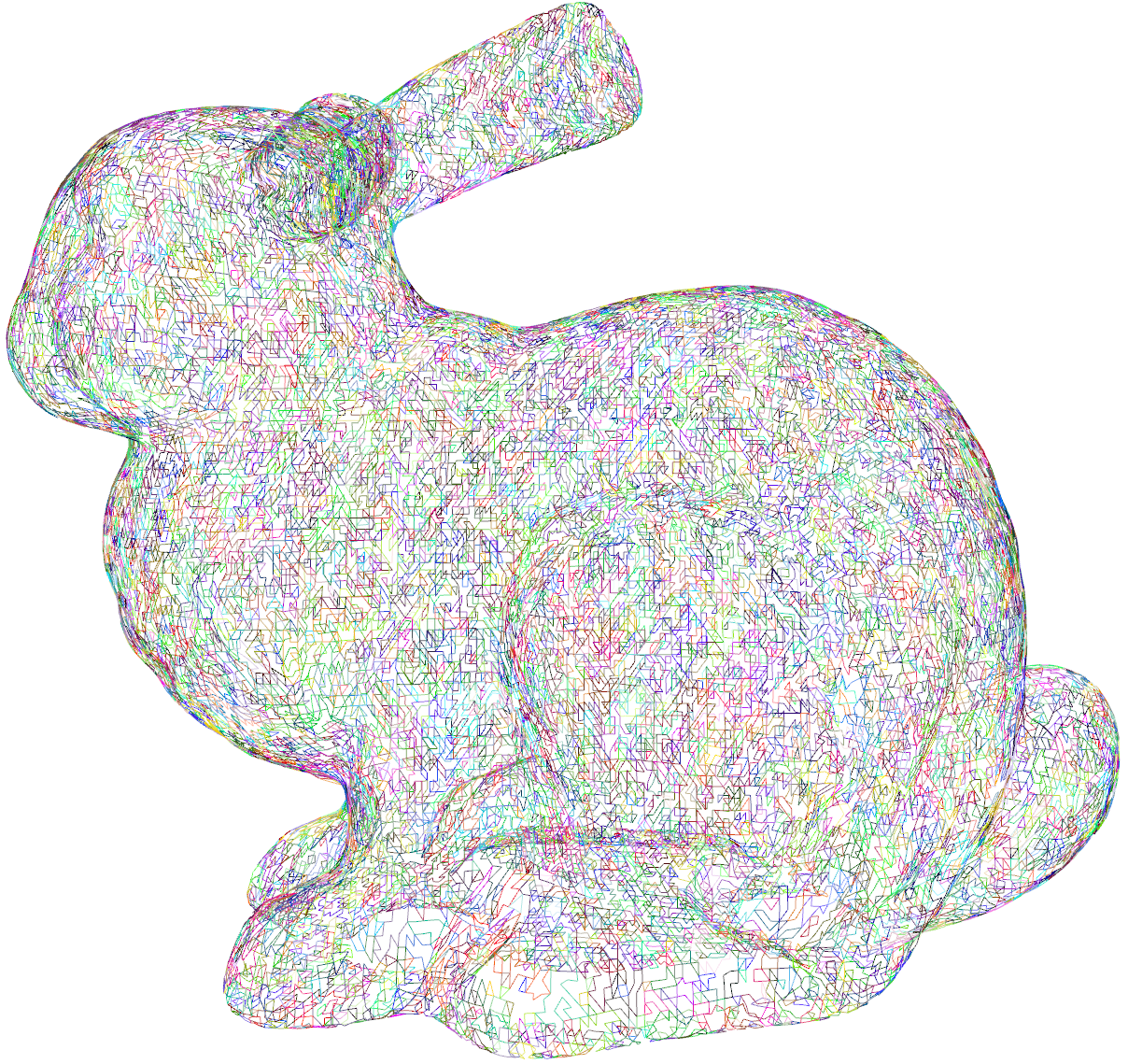}
        \caption{Holes without mesh.} 
    \end{subfigure}
    \caption{All boundaries have been successfully extracted, regardless of the presence of singular vertices. In total, 9724 boundaries were identified, of which 99.9\% contain singular vertices. All half-edges contribute to form the boundary.}
    \label{fig:bunny}
\end{figure}

To demonstrate the robustness of our method against the presence of singular vertices, the original bunny mesh consists of 69451 triangles, we randomly eliminated half of them to introduce half edges, leaving 34725 triangles, see \cref{fig:bunny} for visualization. When our method was applied to this altered mesh, we identified 9724 holes. 
9704 of these boundaries presented singular vertices, accounting for
99.9\% of the detected boundaries. This indicates that using common method \cite{liepa2003filling}, a substantial 
99.9\% of these boundaries would go undetected. Additionally, we ensured that all half-edges were utilized once in the boundaries construction process. Note that on this particular context, making a distinction between coastlines (main boundaries), tide holes, and lake holes becomes irrelevant.

\subsection{A simple triangle mesh}
\begin{figure}[htbp]
   \centering
    \begin{subfigure}[t]{0.45\textwidth}
        \centering
        \includegraphics[width=\textwidth]{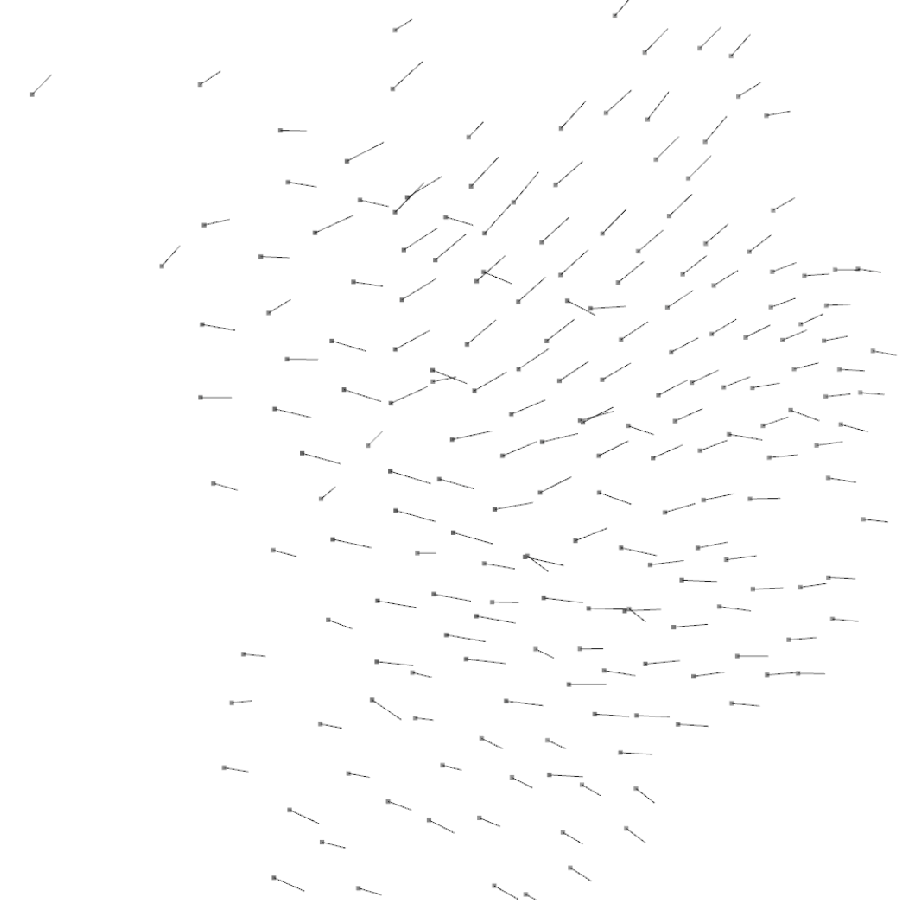}
        \caption{Point cloud with normal vectors.}
        \label{fig:before-malta-recon-a}
    \end{subfigure}%
    ~ 
    \begin{subfigure}[t]{0.45\textwidth}
        \centering
        \includegraphics[width=\textwidth]{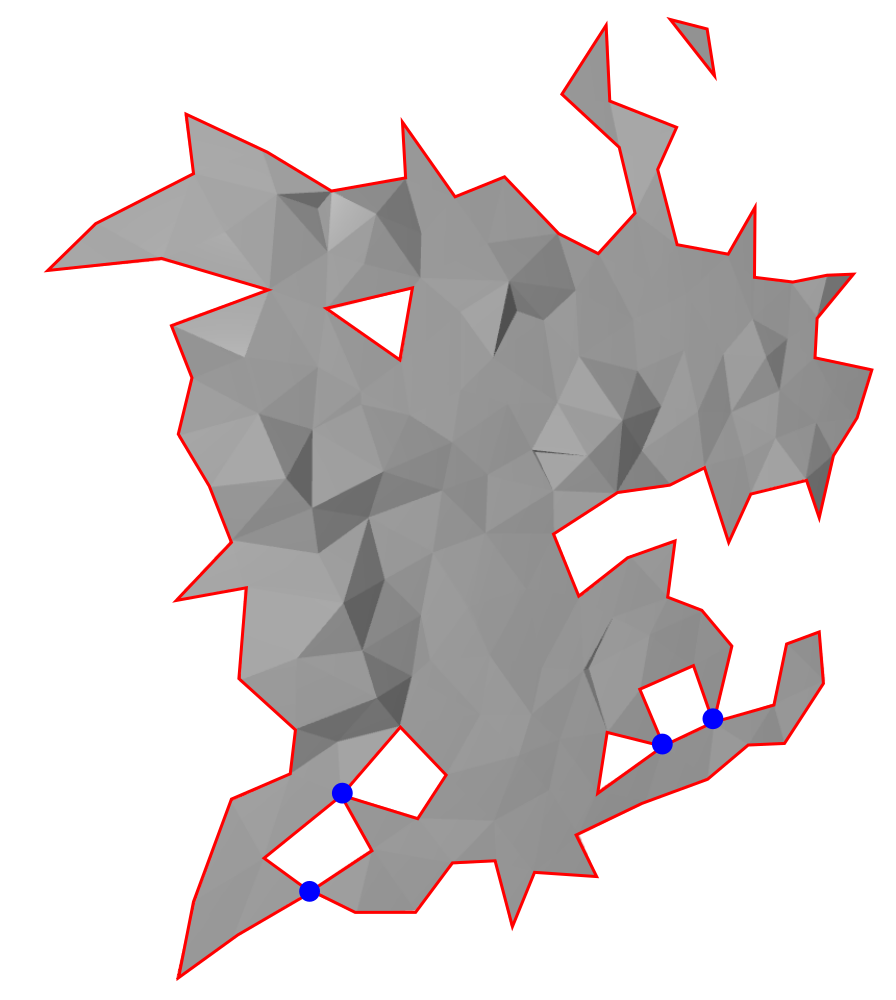}
        \caption{Red line segments indicate half-edges; Blue dots indicate singular vertices.}
        \label{fig:before-malta-recon-b}
    \end{subfigure}%
    \caption{Ball Pivoting Algorithm reconstruction: From point cloud to surface. }
    \label{fig:before-malta-recon}
\end{figure}

\begin{figure}[htbp]
    \begin{subfigure}[t]{0.49\textwidth}
        \centering
        \includegraphics[width=\textwidth]{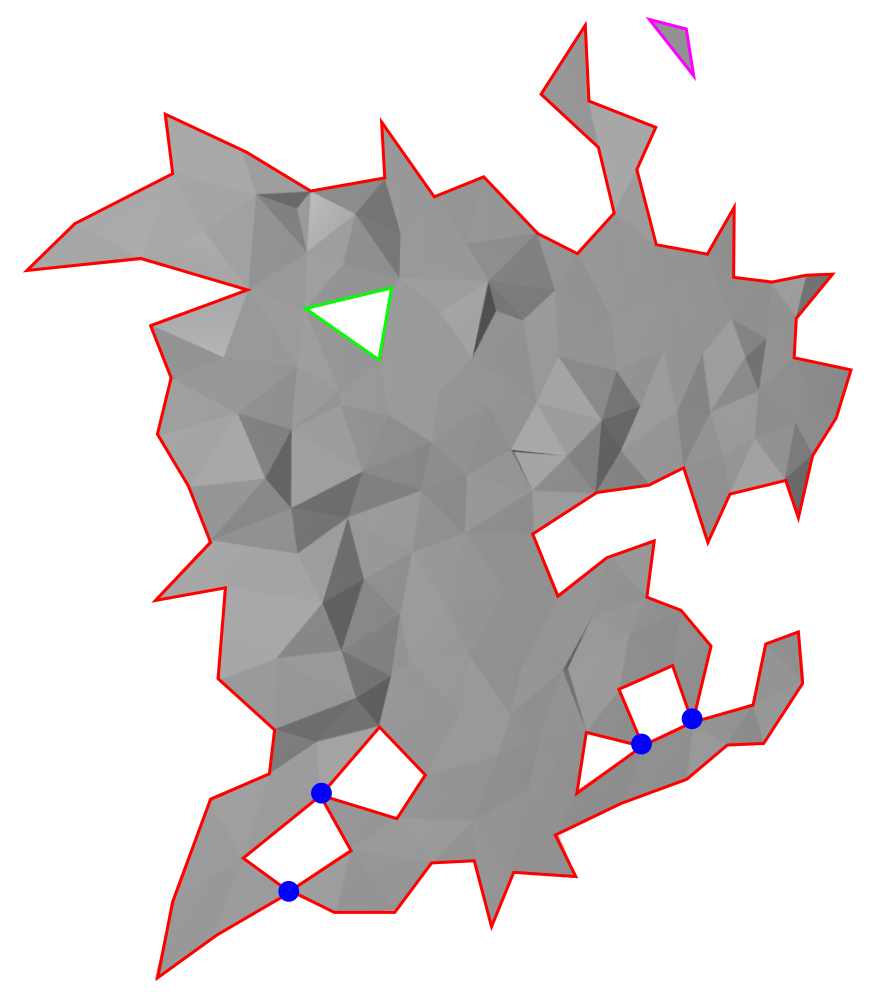}
        \caption{Complex and simple boundaries.}
        \label{fig:before-malta-circuits}
    \end{subfigure}%
    ~ 
    \begin{subfigure}[t]{0.49\textwidth}
        \centering
        \includegraphics[width=\textwidth]{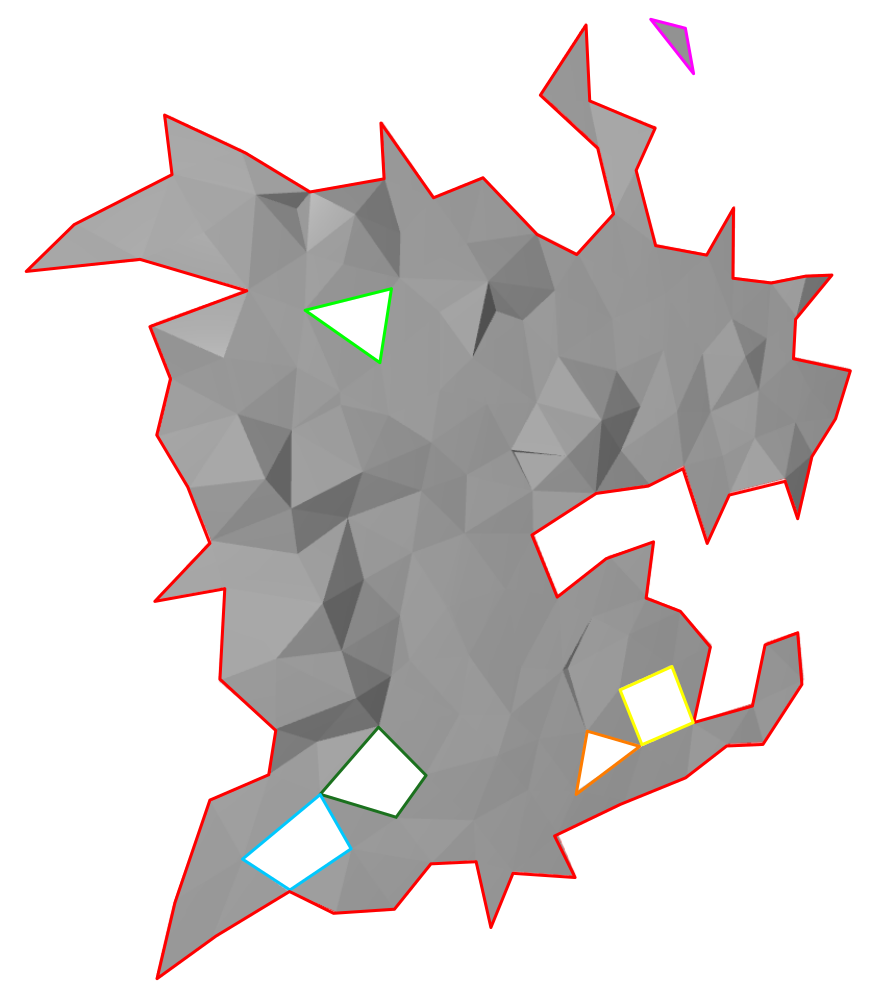}
        \caption{Simple boundaries only.}
        \label{fig:before-malta:circles}
    \end{subfigure}
    \caption{Complex boundary decompositions.}
    \label{fig:before-malta:circir}
\end{figure}

A small point cloud was extracted from a photogrammetry model from \citep{tower_wreck}, as shown in \cref{fig:before-malta-recon-a}. BPA with ball radius $0.4$ meter was applied to the point cloud, resulting in a triangle mesh as shown in \cref{fig:before-malta-recon-b}. Four singular vertices were presented in the triangular mesh. Our boundary reconstruction method was applied to the set of half-edges, resulting in three boundaries (as shown in red, green, and pink in \cref{fig:before-malta-circuits}). Note that the red boundary was a complex boundary.
Our complex boundary decomposition was applied to \cref{fig:before-malta-circuits} and resulted in the red complex boundary being decomposed into several simple boundaries as shown in \cref{fig:before-malta:circles}.

\subsection{A complex triangular mesh}
\begin{figure}[htbp]
    \centering
    \begin{subfigure}[t]{0.490\textwidth}
        \includegraphics[width=\textwidth]{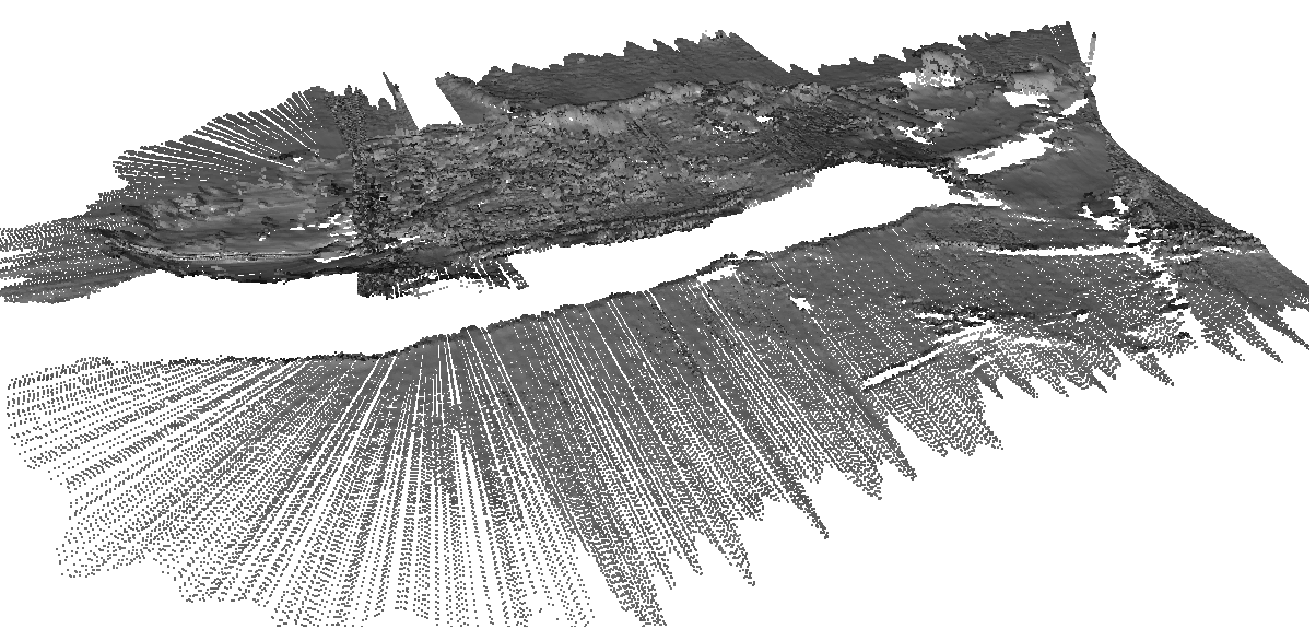}
    \end{subfigure}
    \begin{subfigure}[t]{0.490\textwidth}
        \includegraphics[width=\textwidth]{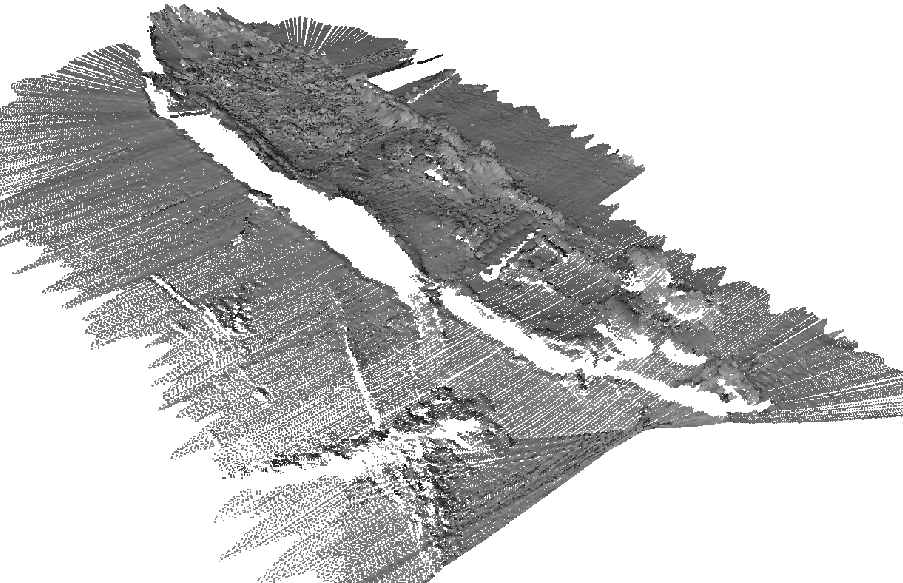}
    \end{subfigure}
    \\  
    \begin{subfigure}[t]{0.490\textwidth}
        \includegraphics[width=\textwidth]{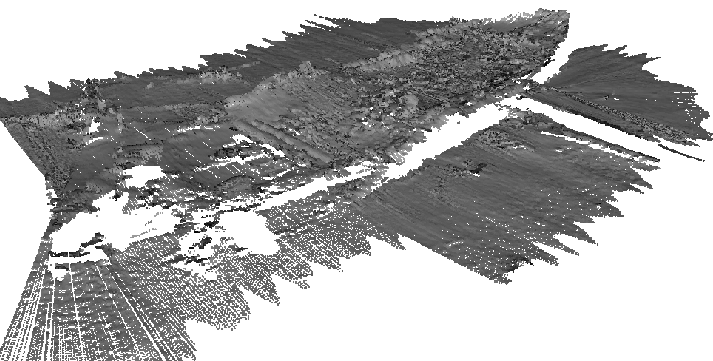}
    \end{subfigure}
    \begin{subfigure}[t]{0.490\textwidth}
        \includegraphics[width=\textwidth]{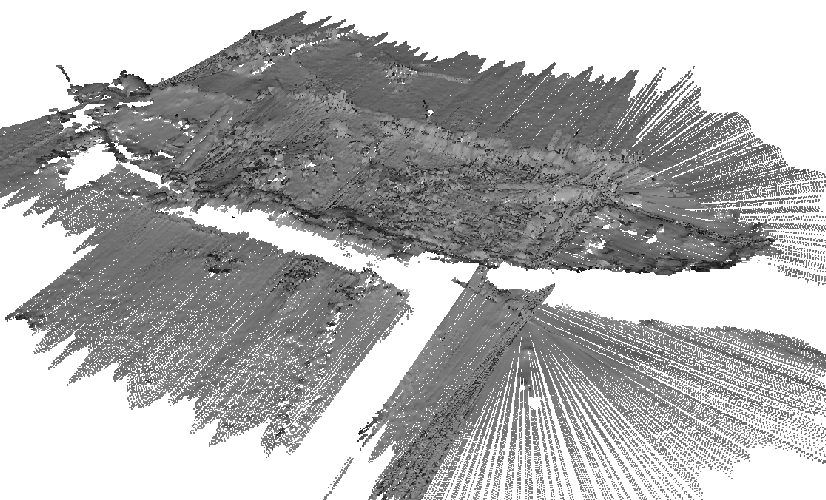}
    \end{subfigure}
    \caption{The point cloud of the Figaro shipwreck (\citet{mogstad2020mapping}), obtained using a MBES, comprises a total of $896,500$ points. For more details on the wreck, see \citep{mogstad2020mapping}. Be aware that their data were collected before March 2020. Our MBES data was gathered in February 2023.}
    \label{fig:figaro_pcd}
\end{figure}

\begin{figure}[htbp]
    \centering
    \begin{subfigure}[t]{0.490\textwidth}
        \includegraphics[width=\textwidth]{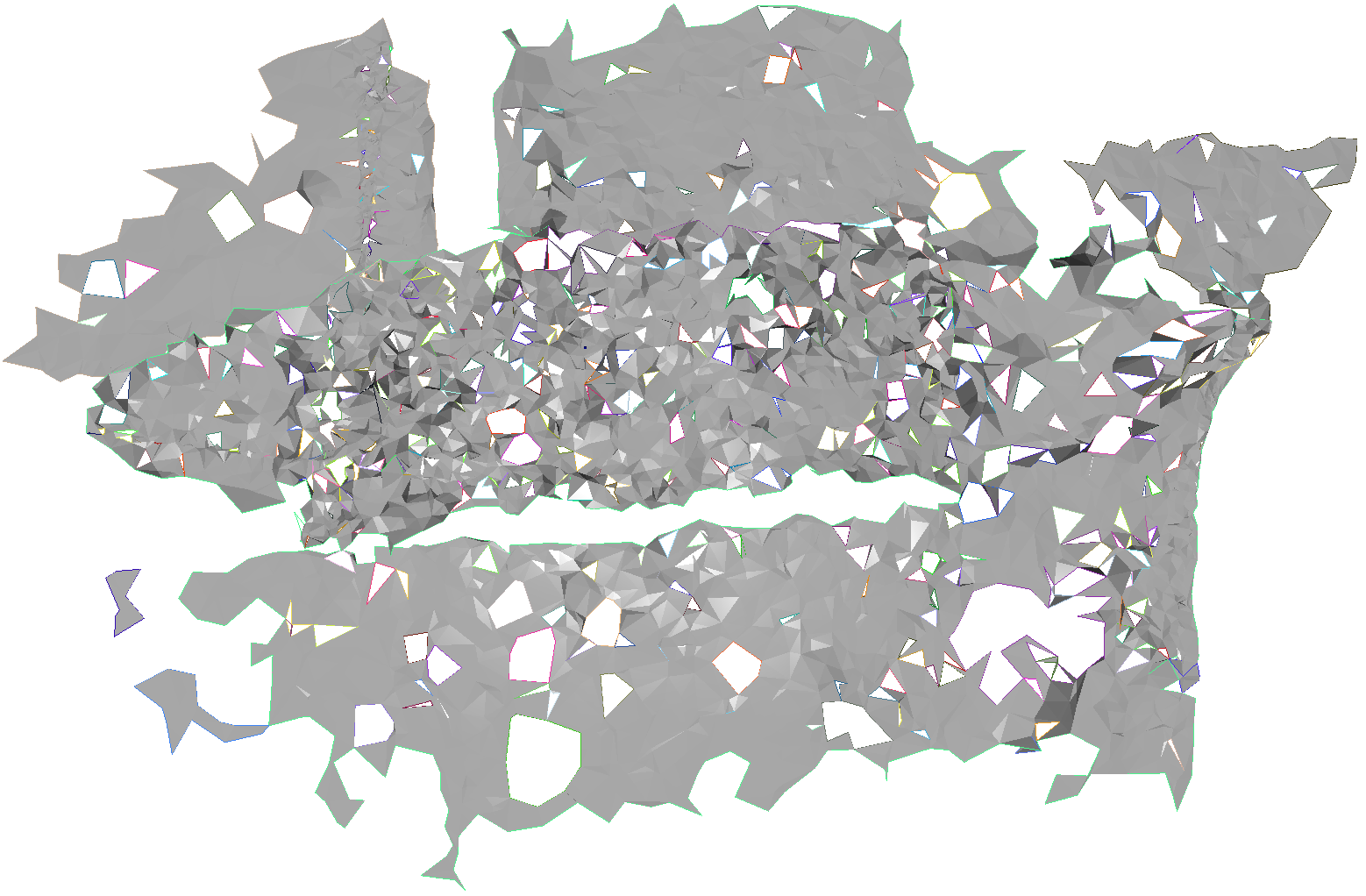}
        \caption{Surface model from BPA.}
        \label{fig:figaro_all:a}
    \end{subfigure}%
    ~
    \begin{subfigure}[t]{0.490\textwidth}
        \includegraphics[width=\textwidth]{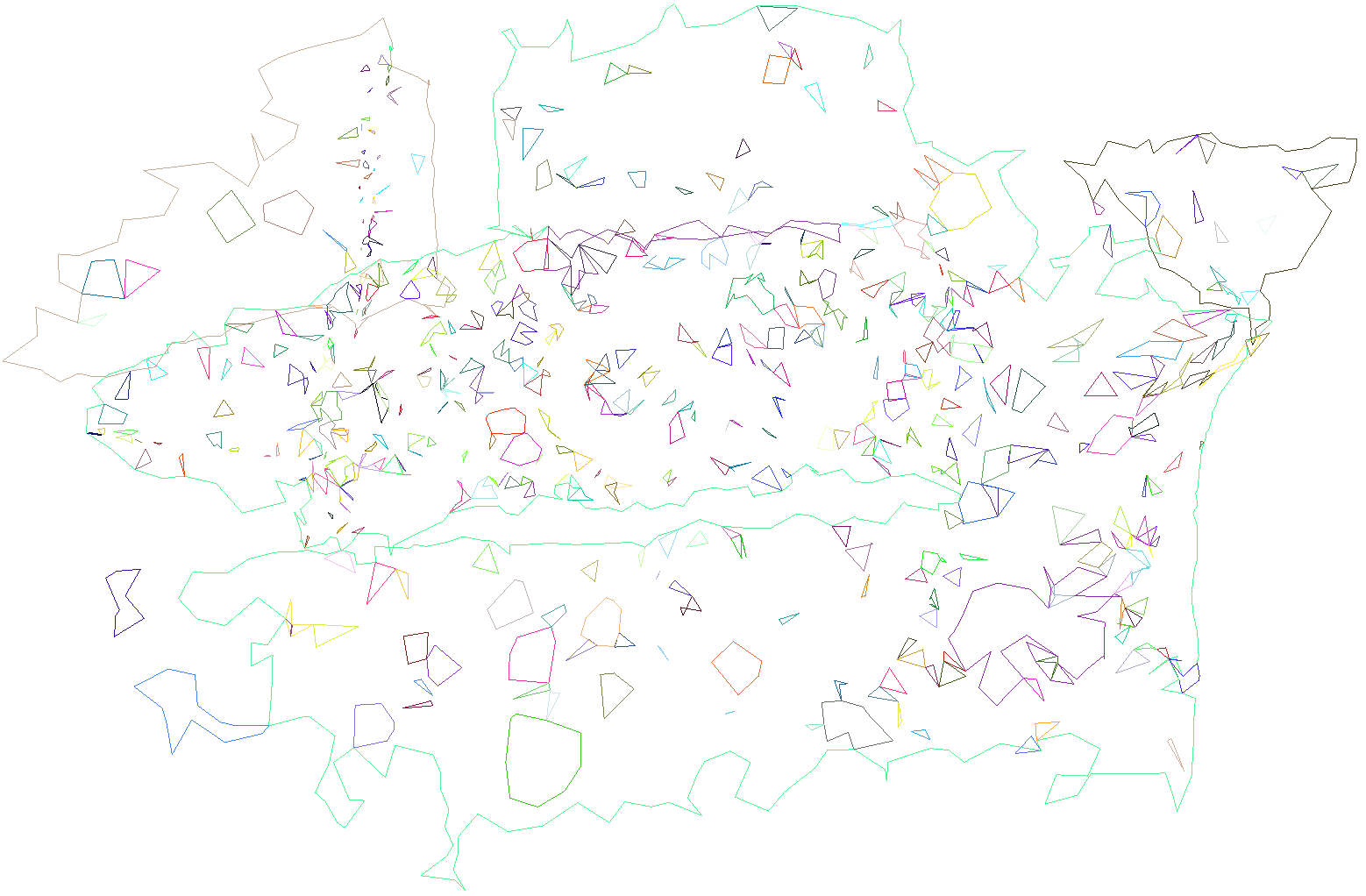}
        \caption{All the detected coastlines and holes.} 
        \label{fig:figaro_all:b}
    \end{subfigure}
    \caption{Surface model and hole detection.}
    \label{fig:figaro_all}
\end{figure}

A Multibeam Echo Sounder (MBES) 
was used to capture a point cloud of the shipwreck Figaro (\citet{mogstad2020mapping}) in Svalbard, Norway. The shipwreck had dimensions of $54 \times 10 \times 6$ meters and sank on 25 July 1908. The MBES was integrated into a snake-like robot (\citet{liljeback2017eelume}), which was used to obtain a point cloud of the Figaro wreck. The resulting point cloud, containing a total of $896500$ points, was acquired within a span of $15$ minutes. The different viewing angles of the point cloud are shown in \cref{fig:figaro_pcd}.
BPA was employed on the Figaro shipwreck point cloud using ball radii of $0.5$ and $0.7$ meters. The original point cloud was densely packed, potentially leading to surface reconstructions being dominated by tiny holes. 
Only 80,000 points were used as input to the Open3D BPA for time efficiency reasons. This is acceptable for our purpose, which is to show how our hole-detection method works. This subset constitutes $0.8\%$ of the original point cloud.
To ensure accurate normal vector estimation, the initial normal vectors were determined by pointing towards the sky. Subsequently, the normal vectors were refined using the functionality provided by Open3D. This allowed for the generation of a reconstructed triangle mesh showing roughly the Figaro shipwreck structure.
The outcome is depicted in \cref{fig:figaro_all:a}. Subsequently, our hole-detection method was applied to the generated triangle mesh, with the results shown in \cref{fig:figaro_all:b}. Despite the intricate nature of the half-edges within the triangle mesh shown in \cref{fig:figaro_all:a}, which includes many singular vertices, our method determined boundaries and identified holes for each half-edge within the triangle mesh. There were $471$ boundaries extracted; of these, $355$ had at least one singular vertex. In other words, if a hole detection method cannot handle singular vertices, it would miss detecting 75\% of the boundaries.

\begin{figure}[htbp]
    \centering
    \begin{subfigure}[t]{0.490\textwidth}
        \includegraphics[width=\textwidth]{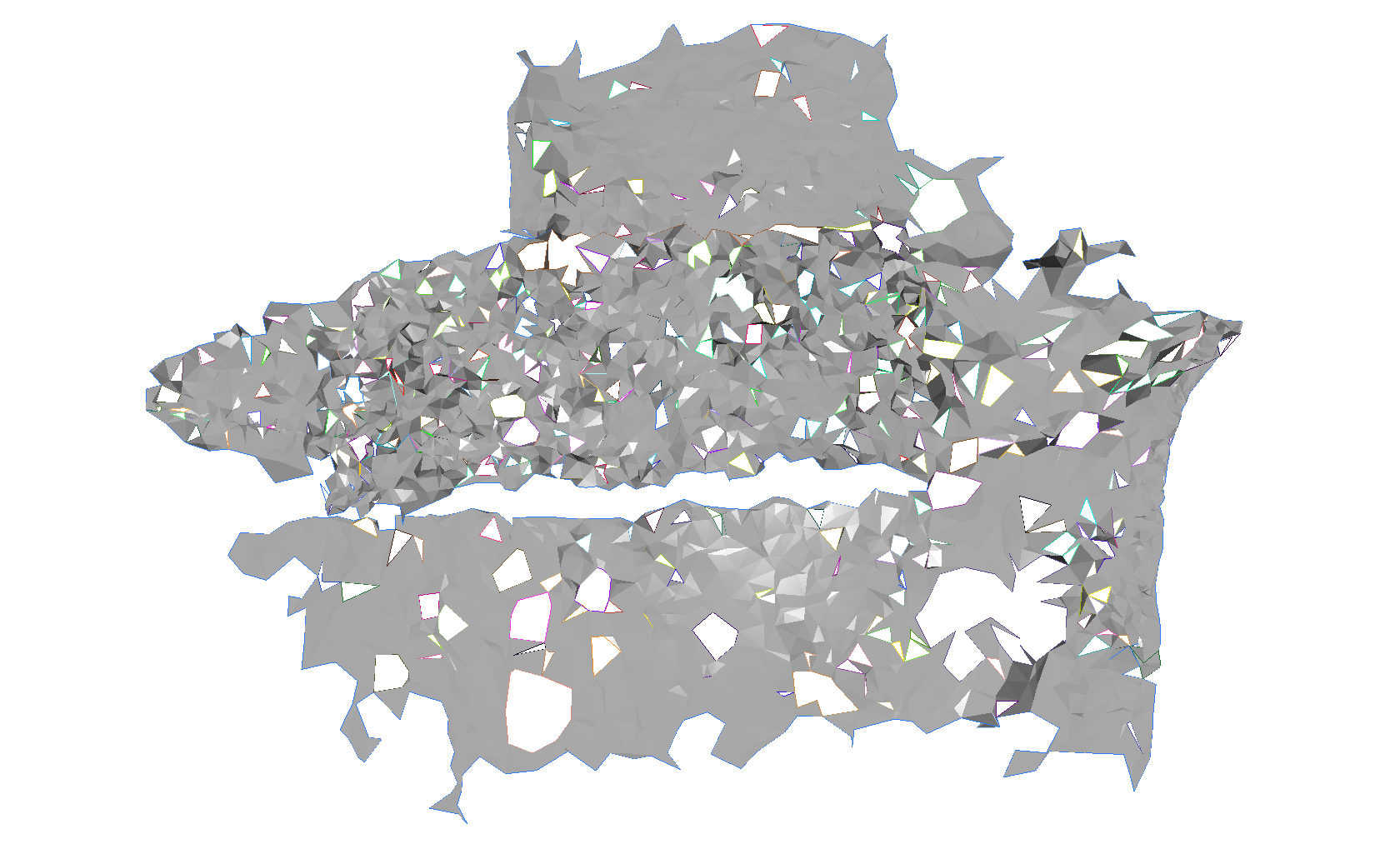}
        \caption{First coastline and its edge-connected mesh.}
        \label{fig:figaro_holes1:a}
    \end{subfigure}
    \begin{subfigure}[t]{0.490\textwidth}
        \includegraphics[width=\textwidth]{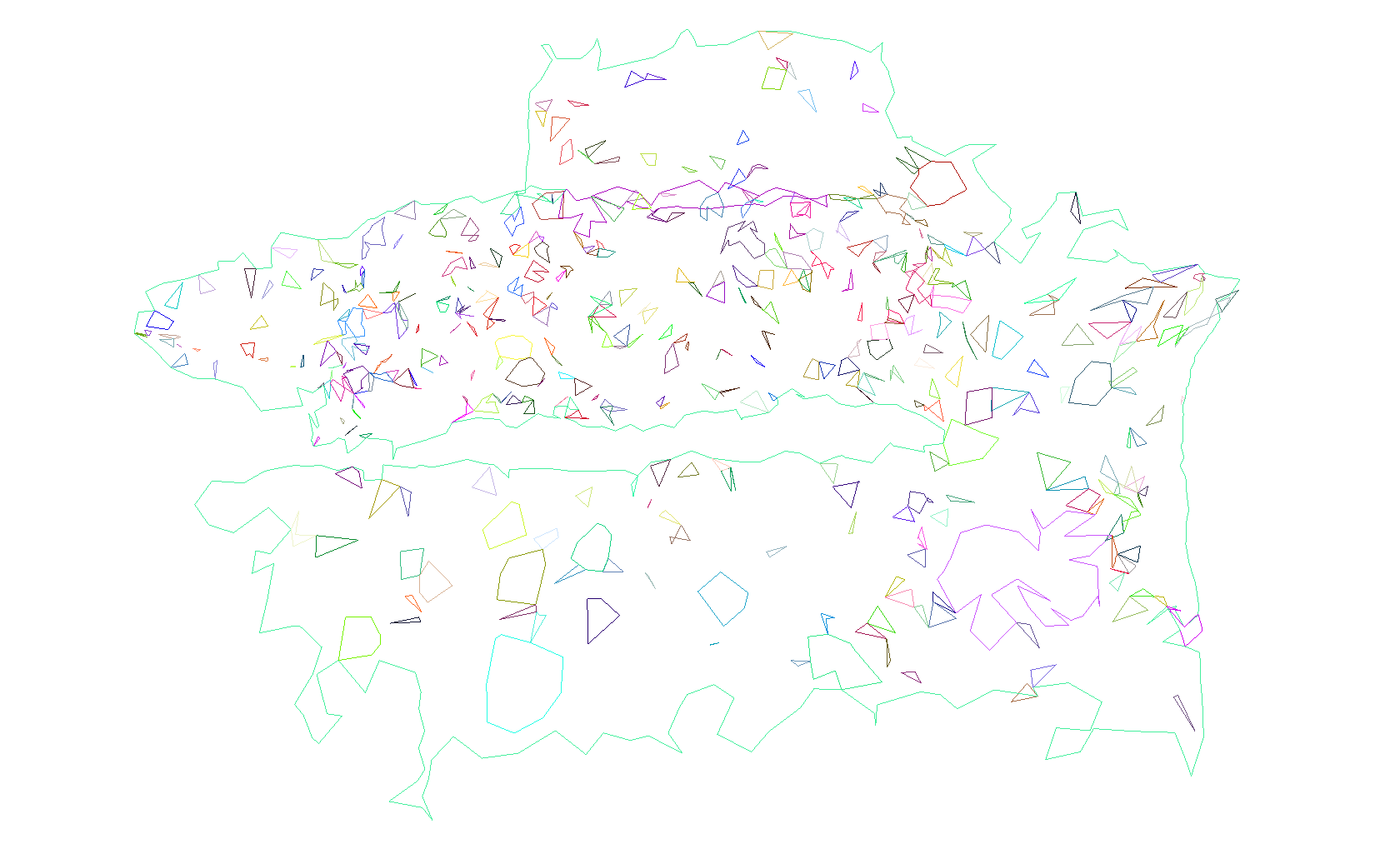}
        \caption{Detected coastline and all holes.} 
        \label{fig:figaro_holes1:b}
    \end{subfigure}
    \\  
    \begin{subfigure}[t]{0.490\textwidth}
        \includegraphics[width=\textwidth]{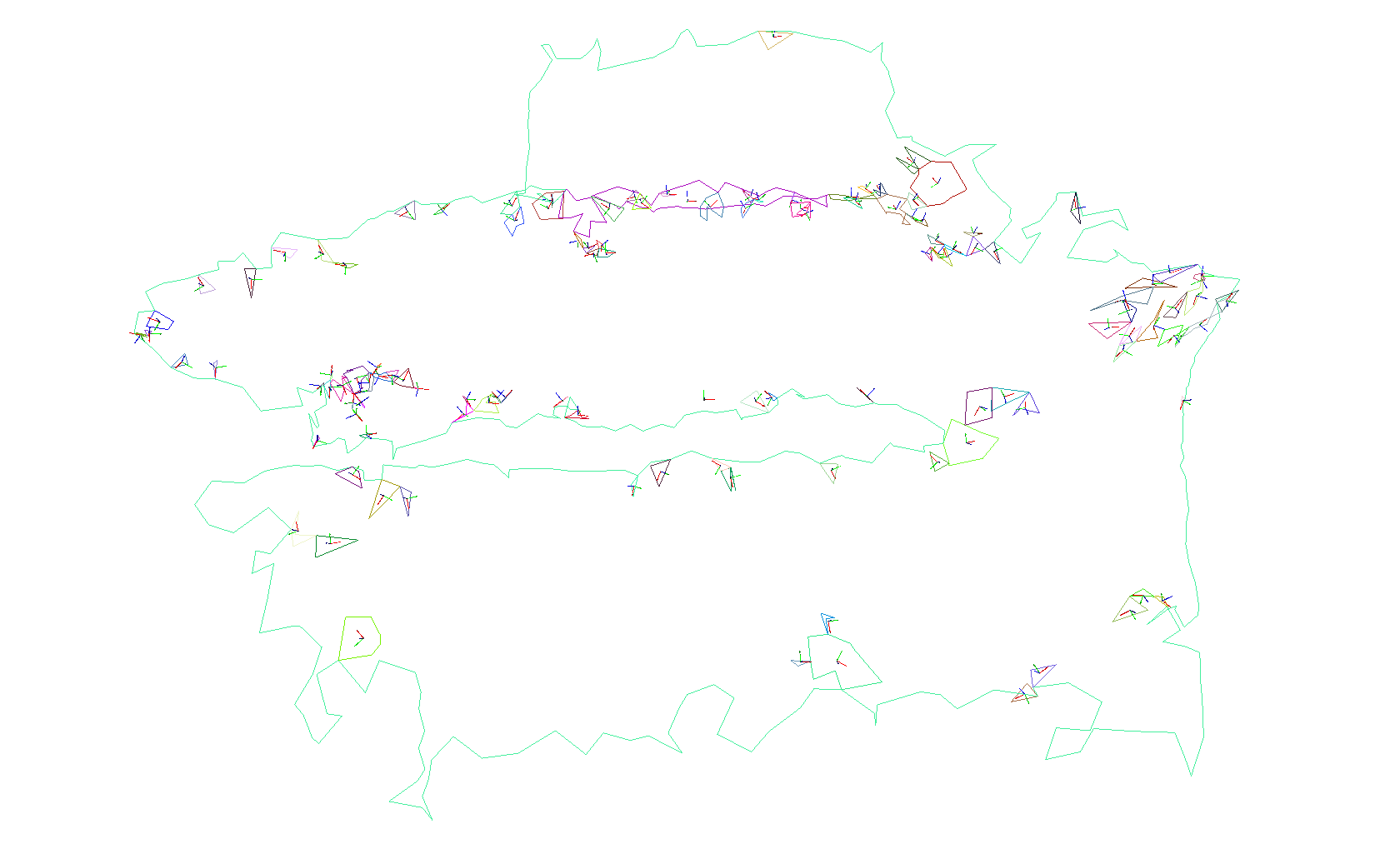}
        \caption{Detected coastline and tide-pool holes.}
        \label{fig:figaro_holes1:c}
    \end{subfigure}
    \begin{subfigure}[t]{0.490\textwidth}
        \includegraphics[width=\textwidth]{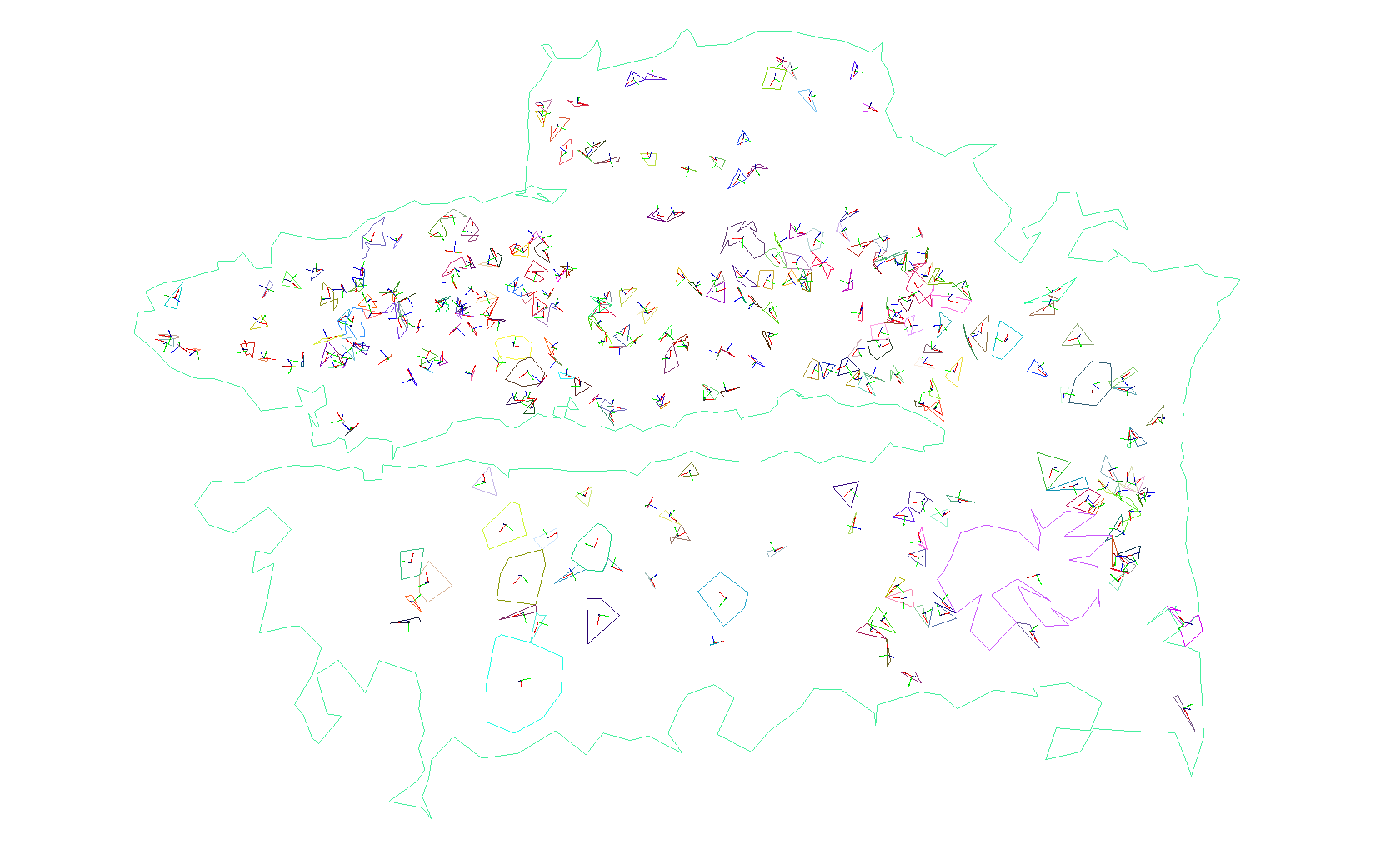}
        \caption{Detected coastline and lake holes.}
        \label{fig:figaro_holes1:d}
    \end{subfigure}
    \caption{Detection of the first coastline and its respective holes.}
    \label{fig:figaro_holes1}
\end{figure}

\begin{figure}[htbp]
    \centering
    \begin{subfigure}[t]{0.490\textwidth}
        \includegraphics[width=\textwidth]{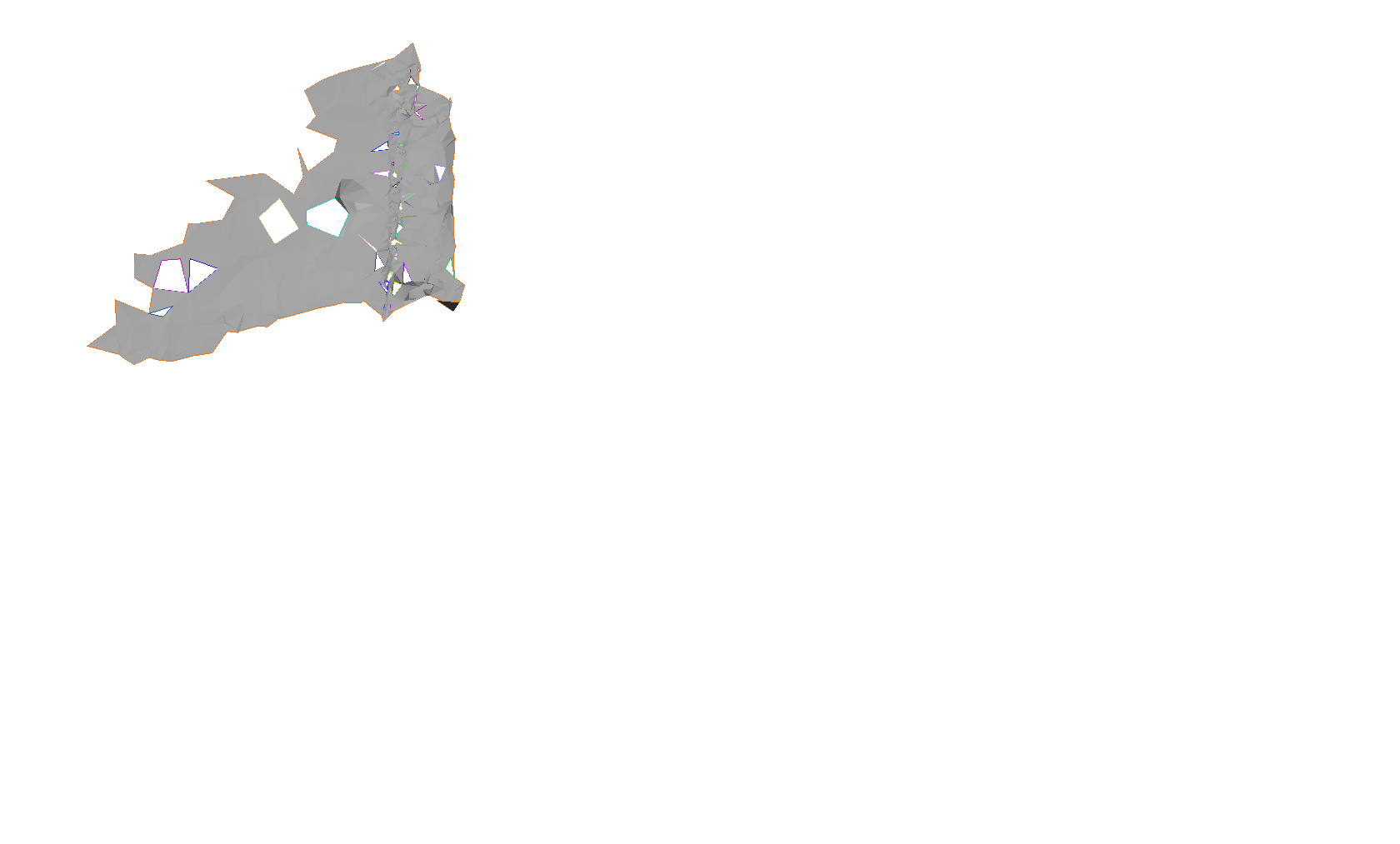}
        \caption{Second coastline and its edge-connected mesh. It has the same camera location as in \cref{fig:figaro_all:a}. A smaller figure is used here to maintain the same camera pose as in \cref{fig:figaro_all:a}, allowing for easier comparison.}
        \label{fig:figaro_holes2:a}
    \end{subfigure}
    \begin{subfigure}[t]{0.490\textwidth}
        \includegraphics[width=\textwidth]{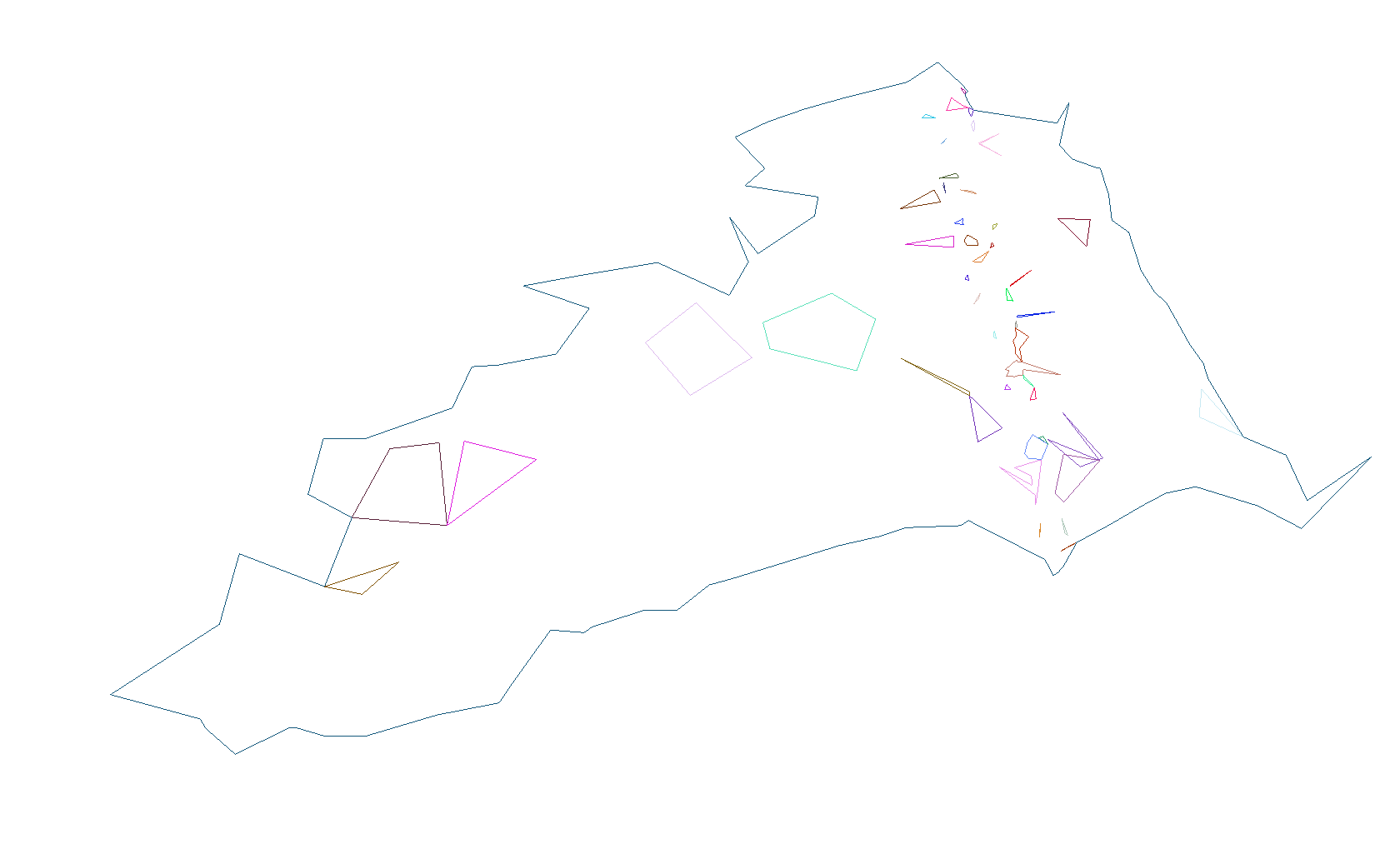}
        \caption{Coastline and all holes (zoomed).} 
        \label{fig:figaro_holes2:b}
    \end{subfigure}
    \\  
    \begin{subfigure}[t]{0.490\textwidth}
        \includegraphics[width=\textwidth]{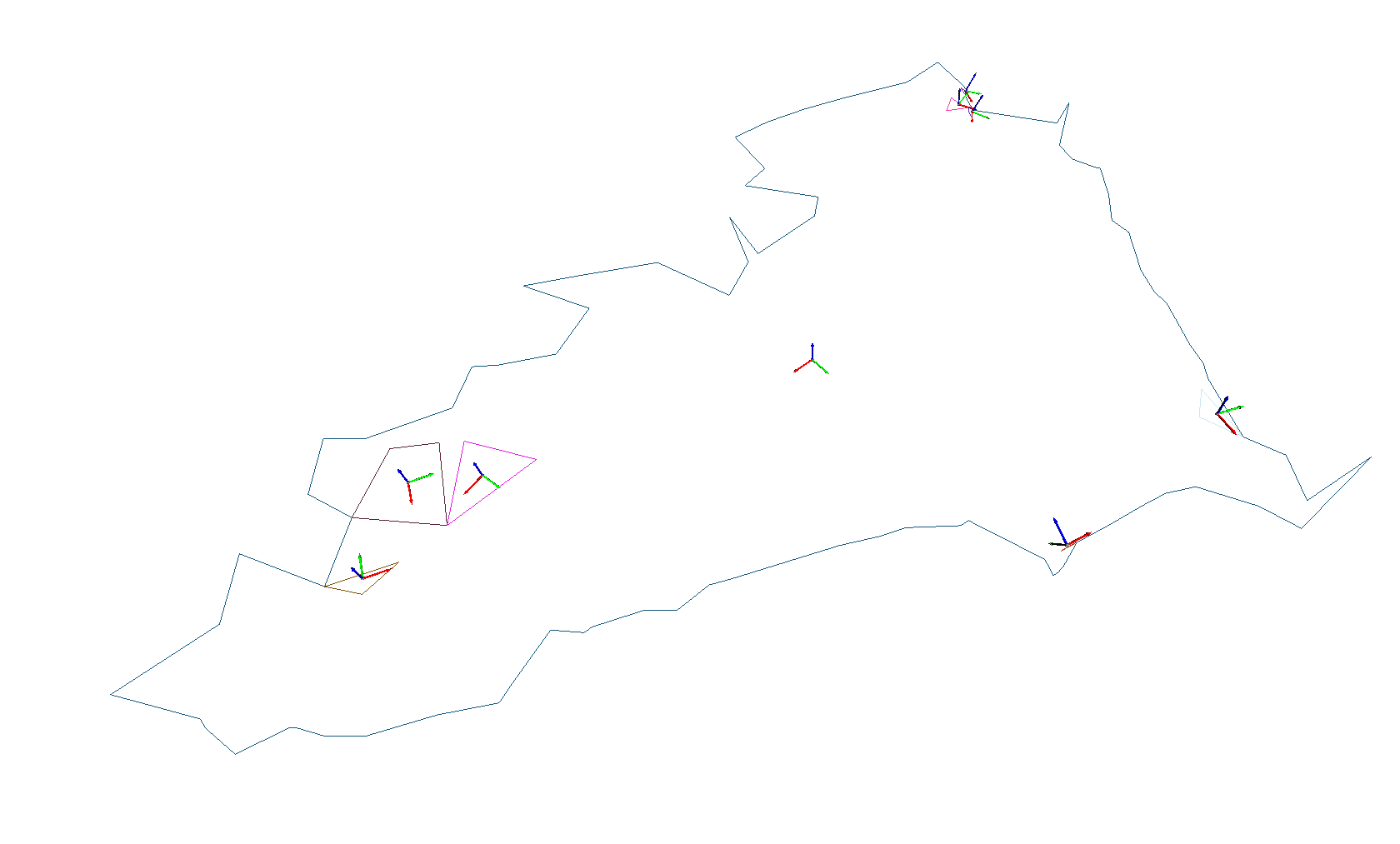}
        \caption{Coastline and tide-pool holes (zoomed).}
        \label{fig:figaro_holes2:c}
    \end{subfigure}
    \begin{subfigure}[t]{0.490\textwidth}
        \includegraphics[width=\textwidth]{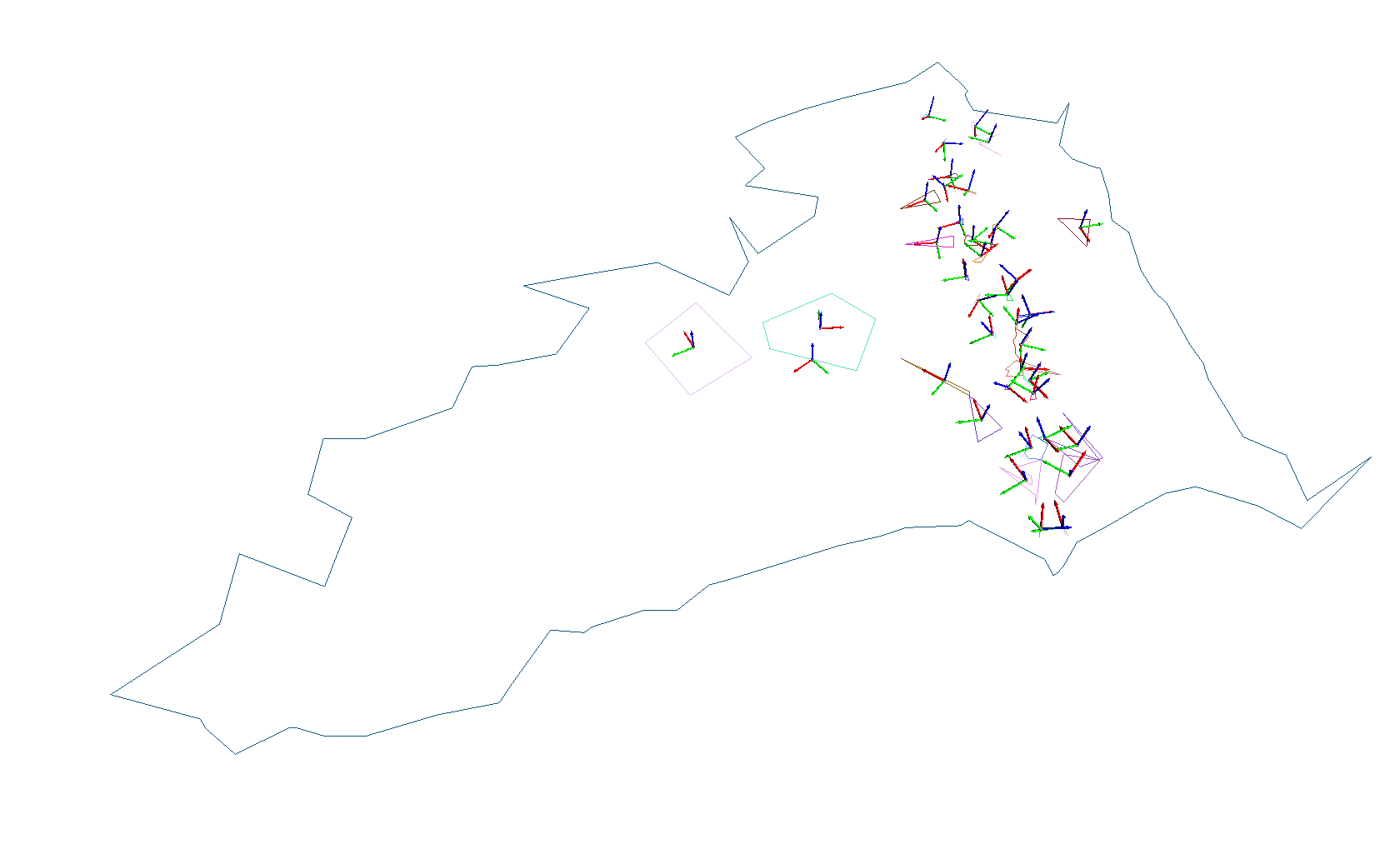}
        \caption{Coastline and lake holes (zoomed).}
        \label{fig:figaro_holes2:d}
    \end{subfigure}
    \caption{Detected second coastline and its holes.}
    \label{fig:figaro_holes2}
\end{figure}

The model contains a total of $16$ coastlines, as depicted in \cref{fig:figaro_all}. The top three coastlines, along with their respective lengths, tide-pool holes, and lake holes, are listed in \cref{tab:holes}. The primary two coastlines are illustrated in \cref{fig:figaro_holes1} and \cref{fig:figaro_holes2}, respectively.
Regarding the coastline depicted in \cref{fig:figaro_holes1}, there are a total of $120$ tide-pool holes (\cref{fig:figaro_holes1:c}) and $276$ lake holes (\cref{fig:figaro_holes1:d}).

\begin{table}[htbp]
\begin{tabular}{|c|c|c|c|}
\hline
Coastline index: & Length& \# Tide holes: & \# Lake holes: \\ \hline
1 & 341.58 meter & 120 & 276 \\ \hline
2 & 80.65 meter & 8 & 38 \\ \hline
3 & 49.03 meter & 5 & 8 \\ \hline
4 -16 & Average 4.33 meter & 0 & 0 \\ \hline
\end{tabular}
\caption{Categorization of the holes in the surface model \cref{fig:figaro_all}.}
\label{tab:holes}
\end{table}

\section{Conclusion} \label{sec:concl}
As discussed in \cref{sec:preli}, the primary challenge lies in establishing boundaries within an edge-manifold triangle mesh when dealing with singular vertices.
Consequently, we have introduced a new and robust technique for identifying boundaries within an edge-manifold triangle mesh, regardless of the presence of singular vertices. We have also supplied two mathematical theorems and their corresponding proofs (in \ref{app:thm}), ensuring the presence of one and only one boundary for every half-edge using our methodology.
In addition, we provided a novel and robust way to decompose a complex boundary/hole (boundary/hole with repeated vertices) into several simple boundaries/holes (boundaries/hole without repeated vertices).
The introduced method is proven to obtain holes robustly in an edge-manifold triangle mesh.
Moreover, we can segment the boundaries into coastlines (main boundaries) and different types of holes. 
We have also tested our hole-detection method on three distinct triangle meshes with holes. It shows that our method can detect and categorize all boundaries into different holes. Significantly, one of the meshes is generated based on real acoustic data, highlighting the practical and real-world applicability of our proposed method.
We provide the source code of our method for hole-detection for the benefit of communities in both CAD and underwater robotics. As of now, our approach relies solely on Python and involves searching for neighboring triangles that span the entire mesh. In future developments, we aim to harness the half-edge data structure to enhance efficiency.

Our contribution involves: 
\begin{itemize}
    \item A method to detect and extract all holes in an edge-manifold mesh without any projection from 3D to 2D, comparing to \citet{gou2022limofilling}.
    \item The only assumption made is that the triangle mesh is an edge-manifold mesh. This assumption is notably less restrictive in comparison to the common assumptions made in related work, which often necessitate the triangle mesh to be oriented, connected, and manifold (inclusive of edge-manifold).
    \item A mathematical theorem (\cref{thm:next-edge}, Appendix) and proof (see \cref{app:thm}) is provided that shows the proposed boundary (hole) detection method can extract boundaries for all half-edges, even in the presence of singular vertices. This implies that no matter how complex the triangle mesh is, we can reliably determine a boundary/hole associated with every half-edge.
    \item A approach is described to decompose complex boundary (boundary with repeated vertices) into simple boundaries (boundaries with no repeated vertices). A mathematical theorem (\cref{thm:decom}, Appendix) and proof (see Appendix) are provided to demonstrate the feasibility of achieving this transformation in all cases.
    \item A method to classify main boundaries (known as model boundaries) and holes from simple boundaries is proposed.
    \item Source code demonstrating the implementation of the proposed method is provided for the benefit of the community. \\ {\color{red} https://github.com/Mauhing/hole-detection-on-triangle-mesh}.
\end{itemize}

\section{Acknowledgement}
This work was supported by the Research Council of Norway (RCN) through the Autonomous Robots for Ocean Sustainability (AROS) project (project number 304667) and the Center of Excellence, NTNU AMOS - Autonomous Marine Operations and Systems  (project number 223254), as well as the NTNU VISTA Centre for Autonomous Robotic Operations Subsea (CAROS).

We would like to thank NTNU AUR-lab, NTNU AMOS, NTNU VISTA CAROS for providing the multibeam echosounder (MBES) data.  A special note of appreciation goes to Dr. Ture Fronczek-Munter from Eelume AS for his assistance in obtaining the MBES data. Additionally, we would like to express our thanks to the research group led by Prof. Timmy Gambin at the University of Malta and Heritage Malta for their contribution of photogrammetry data (supported by the EEA and Norway Grants).

\appendix
\section{Additional definitions, lemma, and theorems}
\label{app:thm}
\begin{definition}[\textbf{Boundary triangle}]
    A boundary triangle is a triangle $t_{ijk}$ that contains at least one half-edge.
\end{definition}

\begin{definition}[\textbf{Transition triangle set}]\label{def:transition-traingle-set}
    Given a triangle mesh $\set T$ and a set $\set H$ that contains all the half-edges of $\set T$, the transition triangle set of $h_{ij} \in \set H$, denoted as $\TTS(h_{ij})$, is defined such that the following hold true:
    \begin{enumerate}
        \item $\TTS(h_{ij}) \subseteq \set R(v_j)\subseteq \set T$. $\set R(v_j)$ is the set of 1-ring triangles of vertex $v_j$ (see \cref{def:one-ring-triangles}).
        \item $t_{ijk} \in \TTS(h_{ij})$. The permutation of $ijk$ does not matter.
        \item $\TTS(h_{ij})$ is an edge-connected mesh (see \cref{def:edge-connected-mesh}).
        \item $\forall t_u \in \set R(v_j)$ that is edge-connected to $\forall t_v \in \TTS(h_{ij})$ such that $u \neq v$ $\implies$ $t_u \in \TTS(h_{ij})$.
    \end{enumerate}
\end{definition}

\begin{figure}[htbp]
    \centering
    \begin{subfigure}[t]{0.450\textwidth}
        \includegraphics[width=\textwidth]{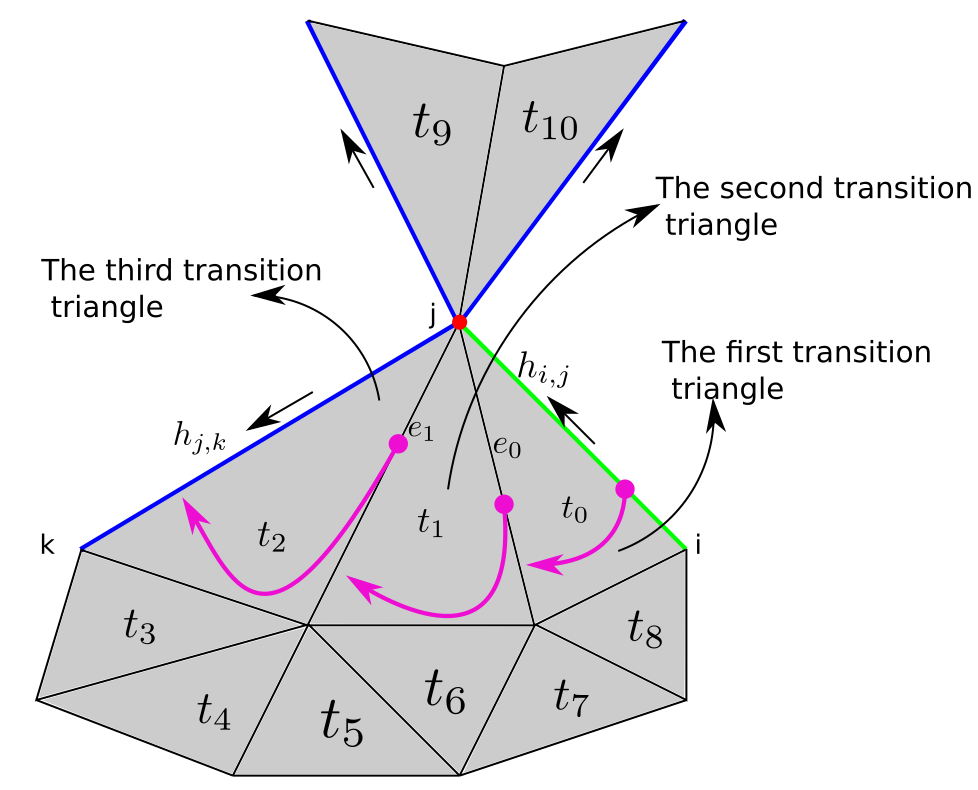}
        \caption{Illustration to explain how to find the next half-edge by using the transition triangle. The subsequent transition triangles are denoted $\TTS(h_{i,j}) = \{t_0, t_1, t_2\}$ in the shown case.}
        \label{fig:lemma:onehalfedge:a}
    \end{subfigure}
    \,
    \begin{subfigure}[t]{0.450\textwidth}
        \includegraphics[width=\textwidth]{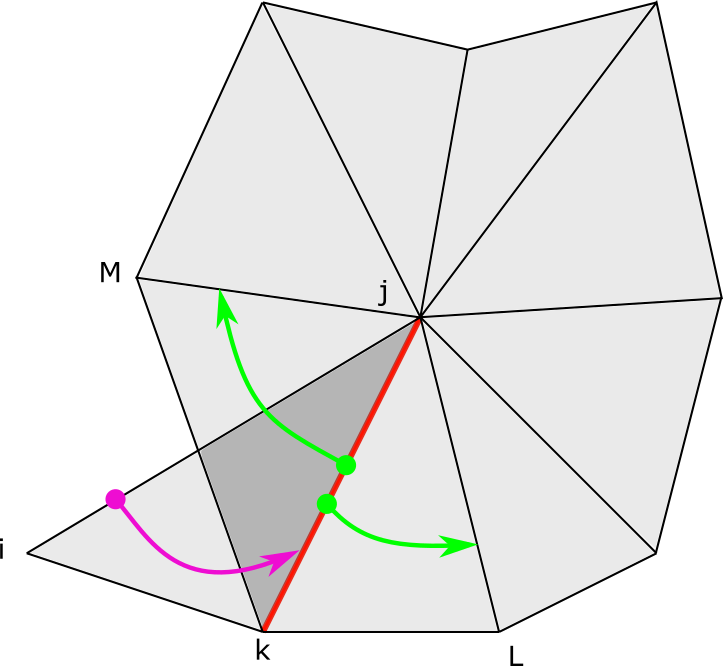}
        \caption{The red edge is a non-manifold edge, implying more than two triangles are adjacent to the red edge. The green curved arrow indicates two other triangles with $e_{jk}$. This mesh is not an edge-manifold mesh.} 
        \label{fig:lemma:onehalfedge:b}
    \end{subfigure}
    \caption{Illustration to explain \cref{lem:onehalfedge}.}
    \label{fig:lemma:onehalfedge}
\end{figure}

Example: In \cref{fig:lemma:onehalfedge:a}, those triangles with the purple curved arrows form the transition triangles set of $h_{i,j}$, $\TTS(h_{i,j})$. The set $\set R(v_j) = \{t_0, t_1, t_2, t_9, t_{10}\} $ can not be $\TTS(h_{i,j})$ otherwise $\TTS(h_{i,j})$ will no longer be a edge-connected mesh. The set $\{t_0, t_1\}$ can not be $\TTS(h_{i,j})$ because $t_2 \in \set R(v_j)$ is edge-connected to $t_1$, but $t_2 \notin \TTS(h_{i,j})$.

\begin{remark}
    $\TTS (h_{ij}) \neq \TTS (h_{ji})$ in general. The direction of $h$, indicated by the two vertices in $h$ matters.
\end{remark}

\begin{lemma}  
    \label{lem:onehalfedge}
    If $\set T$ is an edge-manifold triangle mesh, and $\set H$ is the set that contains all the half-edges of $\set T$, for any half-edge $h_{i,j}\in \set H$, there exists one and only one half-edge  $h_{j,k}$, in $\TTS(h_{i,j})$ such that it contains vertex $v_j$ but not vertex $v_i$.
\end{lemma}
\begin{proof}
    $ $\newline
    \textbf{Existence}:
    In this proof, our aim is to show the existence of $h_{j,k}$ by using the procedure defined by \Cref{alg:next-halfedge} to construct $\TTS(h_{ij})$. First, we initialized an empty array and denoted it as $\set W$. Let us assume we have the 1-ring triangles set $\set R(v_j)$. We use $\set W$ to collect triangles $t \in \set R(v_j)$ and show $\TTS(h_{ij}) = \set W$.
    Let us define $e_{-1}:= h_{ij}$
    Starting with $e_{-1}$, there exists one and only one boundary triangle, $t_0$, by \cref{def:half-edge}. $t_0$ is inserted into $\set W$. 
    We find the transition edge, $e_0$, of $e_{-1}$ with $t_0$.  
    Edge $e_0$ can either be a half-edge or a full-edge due to $\set T$ being an edge-manifold mesh.
    If $e_0$ is a full-edge, we jump to Case A with $n=0$. If $e_0$ is a half-edge, we jump to Case B with $n=0$

    \paragraph{Case A: full-edge}
    In the case of $e_n \notin \set H$, $e_n$ is a full edge, which is an edge adjacent to two different triangles. There exists one and only one triangle $t_{n+1}$ that has an edge $e_n$ but not $e_{n-1}$ since $\set T$ is edge-manifold. Since $t_{n+1}$ has edge $e_n$, $t_{n+1}$ has vertex $v_j$. This implies $t_{n+1} \in \set R(v_j)$, see \cref{fig:lemma:onehalfedge:a} for illustration. $t_{n+1}$ is inserted into $\set W$.
    (If $\set T$ is not edge-manifold, there could be more than one triangle that has an edge $e_{n}$ but not $e_{n-1}$, see \cref{fig:lemma:onehalfedge:b} for illustration).
    We find the transition edge $e_{n+1}$ of $e_n$ with $t_{n+1}$.
    If $e_{n+1}$ is a full-edge, we jump to Case A with $n:= n+1$. If $e_{n+1}$ is a half-edge, we jump to Case B with $n:= n+1$.

    \paragraph{Case B: half-edge}
    In the case of $e_n \in \set H$, we first want to show $\TTS(h_{ij}) = \set W$, then $h_{j,k} = e_n$.
    \begin{enumerate}
        \item $\forall t \in \set W$, $t$ has vertex $v_j$. Therefore, $ \set W \in \set R(v_j)$
        \item $t_0 \in \set R(v_j)$ where $t_0$ has vertice $v_i$ and $v_j$.
        \item For $l > 0$, every $t_l$ triangle get inserted into $\set W$, it has to be edge connected to previous triangle $t_{l-1}$. For $l = 0$, we have $t_0 \in \set R(v_j)$ already. This implies $\set W$ is an edge-connected mesh.
        \item Assuming there $t_u \in \set R(v_j)$ that is edge-connected to $\forall t_v \in \set W$ such that $u \neq v$ but $t_u \notin \set W$. $t_u$ can not be $t_0$ since $t_0 \in \set W$. Since $t_u$ is edge-connected to $t_v$, there exists an edge $e_{v}$ shared between $t_u$ and $t_v$. $t_v$ has also edge $e_{v-1}$ that does not belong to $t_u$. However, $t_u$ will be collected in Case if $t_u$ has $e_v$ but not $e_{v-1}$. It contradicts to our assumption. Therefore, $\forall t_u \in \set R(v_j)$ that is edge-connected to $\forall t_v \in \set W$ such that $u \neq v$ $\implies$ $t_u \in \set W$.
    \end{enumerate}
    $\set W$ satisfies \cref{def:transition-traingle-set} $\implies \TTS(h_{ij}) = \set W.$ $e_n$ has vertex $v_j$ because all transition edge has vertex $v_j$. $e_n$ can not have vertex $v_j$ since $e_{-1}$ is a half-edge and has both $v_i$ and $v_j$. If $e_n$ has $v_j$, $e_{-1}$ will not be a half-edge in the first place. Therefore, the existence of the next connected half-edge $h_{j,k}$is guaranteed. 
    
    \textbf{Uniqueness}:
    The aforementioned process uniquely identifies all transition edges and transition triangles. Since $h_{j,k}$ is obtained by traversing all transition edges and transition triangles, the determined half-edge $h_{j,k}$ is unique. 
\end{proof}

\begin{remark}
    In \cref{lem:onehalfedge}, no assumption was made as to whether the vertex $v_{j}$ in $h_{i,j}$ is singular or not.
\end{remark}

\begin{definition}[Upcoming-edge] \label{def:upcoming-edge}
    From \cref{lem:onehalfedge}, for every $h_{i,j} \in \set H$, 
    there exists one and only one half-edge $h_{j,k}$ in the transition triangles $\TTS(h_{j,k})$ such that $h_{j,k}$ has starting vertex $v_j$ and $h_{i,j}$ has ending vertex $v_j$. Then we refer to $h_{j,k}$ as the upcoming-edge of $h_{i,j}$.
\end{definition}

\begin{lemma}
    \label{lem:bijective}
    Let $\set H$ be the set of half-edges of an edge-manifold triangle mesh.
    If $f: \set H \rightarrow \set H$ be a function that maps half-edge $h_{ij}$ to its upcoming-edge $h_{j,k}$. We can express this as $f(h_{i,j})=h_{j,k}$ and \cref{lem:onehalfedge} means that the upcoming-edge exists and can be found. The mapping of $f$ is bijective (one-to-one and on-to).
\end{lemma}
\begin{proof}
    We first prove the one-to-one property: Let $r$ be the function that flips the direction of a half-edge, which means $r(h_{i,j}) = h_{j, i}$.
    Let $f^{-1}: \set H \rightarrow \set H$ be $f^{-1} = r\circ f \circ r $. Given $h_{j,k} = f(h_{i,j})$, we have:
    \begin{align*}     
        f^{-1}(h_{j,k}) &= r\circ f \circ r (h_{j,k}) \\ 
        &= r\circ f (h_{k,j})\\ 
        &= r(h_{j,i})\\ 
        &= h_{i,j}.
    \end{align*}
    \cref{lem:onehalfedge} was applied to obtain $f(h_{k,j}) = h_{j,i}$ by changing index. This shows that the inverse function $f^{-1}$ exists and proves the one-to-one property.
    Now, we prove that $f$ is also on-to: Since the co-domain of $f$ and the domain of $f^{-1}$ are both $\set H$, the co-domain, and range of $f$ are the same. This implies that $f$ is on-to. 
    Since $f$ is one-to-one (injective) and on-to (subjective), $f$ is a bijective function.
\end{proof}

\begin{theorem}
    \label{thm:next-edge}
    If $\set T$ is an edge-manifold triangle mesh, and $\set H$ is the set that contains all the half-edges of $\set T$, there exists a set of boundaries, denoted $\set B$ such that for any half-edges $h \in \set H$, there is one and only one boundary $\set b \in \set B$ with $h \in \set b$.
\end{theorem}
\begin{proof}
$ $\newline
\textbf{Existence}:
Given an arbitrary half-edge $h \in \set H$, we can denote it as starting half-edge $h_0$ . 
From \cref{lem:onehalfedge}, there exists one and only one half-edge $h_1$ as the upcoming-edge (\cref{def:upcoming-edge}) of $h_0$. Applying \cref{lem:onehalfedge} iteratively, $h_{k+1}:=f(h_{k})$, to obtain consecutively connected half-edge $[h_0, h_1, h_2, ......, h_{n-1}]$ and stopping once the next half-edge, denoted as $h_{n}$, is found in previously connected half-edges, $[h_0, h_1, h_2, ......, h_{n-1}]$.
This implies that the consecutively connected half-edges must have a repeated edge or have infinitely many unique half-edges ($n=\infty$). However, since $|\set T|$ is finite ($|\set T| < \infty$ ), this implies that $|\set H|$ is also finite. Since $|\set H|$ is finite, having infinitely many unique half-edges is impossible.

We prove that $h_n = h_0$, which is the starting half-edge, by contradiction.
Assuming that $h_0 \neq h_n$, which means that there exists $h_j$ such that $h_j = h_n$, where $ 0 < j < n$.
In the case of $j = n-1$, this implies $h_j = h_{n-1}$, which implies $h_{n-1} = h_{n}$. This must be false due to \cref{lem:onehalfedge}.
In the case of $ 0 < j < n-1$, both $f(h_{n-1}) = f(h_{j-1}) = h_{j} = h_{n}$, where $n \neq j$. This must be false since the mapping $f$ is a bijective function proven in \cref{lem:bijective}. The only option left is $j=0$. Therefore, $h_n = h_0$ must be true. 
When $h_n = h_0$, $[h_0, h_1, h_2, ......, h_{n-1}]$ forms a boundary by definition.

\textbf{Uniqueness:}
Assuming two boundaries $\set q \neq \set p$, where $\set q \in \set B$, $\set p \in \set B$, and $h \in \set q$, $h \in \set p$. We denote the boundary $\set q :[h, q_1, ..., q_{n-1}]$ and $\set p := [h, p_1, ..., p_{m-1}]$, where $n$ and $m$ are the number of connected half-edges in the set $\set q$, $\set p$ respectively. 
We use the function $f$ from \cref{lem:bijective}, $q_1 = f(h)$ and $p_1 = f(h)$. Since $f: \set H \rightarrow \set H$ is bijective, this implies $q_1 = p_1$. Use \cref{lem:bijective}, iteratively this implies $q_i = p_i$ and $n=m$. And this implies $\set q = \set p$ which contradicts the assumption $\set q \neq \set p$. Therefore, there is only one boundary $\set b$ that contains half-edge $h$.
\end{proof}

\begin{theorem}[Complex boundary decomposition]
    \label{thm:decom}
    If $\set b$ is a complex boundary, then there exist two boundaries $\set b_1$ and $\set b_2$ such that 
    \begin{enumerate}
        \item For every half-edge $h \in \set b$, the half-edge $h$ must belong to either $\set b_1$ or to $\set b_2$, but not to both.
        \item The number of half-edges in $\set b_1$ plus the number of half-edges in $\set b_2$ is equal to the number of half-edges $\set b$. 
        \item Neither $\set b_1$ or $\set b_2$ is identical to $\set b$.
    \end{enumerate}
\end{theorem}
\begin{proof}
$ $\newline
   Since $\set b$ is a complex boundary, there exists at least one repeated vertex $v_j$ such that $\set b = \langle v_0, ..., v_j, ..., v_j, ..., v_n \rangle$ by definition (\cref{def:com_bound}). $\set b$ can also be represented by half-edges with an ordered array, which is 
   \[\set b = [{\color{red}h_{0,1}, ...,h_{i,j}}, {\color{blue}h_{j,k}, ...,h_{q,j}},{\color{green}h_{j,r}, ..., h_{n,0}}].\] The change of color indicates the first two crossings of repeated vertex $v_j$. The first crossing is from $h_{i,j}$ to $h_{j,k}$, the second crossing is from $h_{q,j}$ to $h_{j,r}$.
   We use the first two crossings of vertex $j$ to split up the ordered array $\set b$ into $\set b_1 = [{\color{red}h_{0,1}, ...,h_{i,j}},{\color{green}h_{j,r}, ..., h_{n,0}}]$ and $\set b_2 = [{\color{blue}h_{j,k}, ..., h_{q,j}}]$.
   $\set b_1$ is a boundary because both ${\color{red}h_{0,1}, ...,h_{i,j}}$, ${\color{green}h_{j,r}, ..., h_{n,0}}$ are connected half-edges; $h_{i,j}$ and $h_{j,r}$ can also be connected and form a loop. 
   $\set b_2$ is a boundary because ${\color{blue}h_{j,k}, ..., h_{q,j}}$ are connected half-edges that form a loop.
   
   We now prove \cref{thm:decom} point (1). All half-edges in $\set b$ are unique and split $\set b$ into two boundaries without duplicating any half-edge from the aforementioned procedure. Therefore, for every half-edge $h \in \set b$, the half-edge $h$ must exist in $\set b_1$ or $\set b_2$, but not both.

    We now prove \cref{thm:decom} point (2), let $m$, $m_1$ and $m_2$ be the number of half-edges in $\set b$, $\set b_1$, and $\set b_2$ respectively. $m = m_1 + m_2$ since no half-edge is discarded or duplicated from the aforementioned procedure.

    We now prove \cref{thm:decom} point (3). The smallest number of half-edges required to form a boundary is \(3\). Therefore, both \( \set b_1 \) and \( \set b_2 \) must have at least \(3\) half-edges, resulting in \( m_1 \geq 3 \) and \( m_2 \geq 3 \). Given \( m = m_1 + m_2 \), and  \( m_1 \geq 3 \) and \( m_2 \geq 3 \), neither \( \set b_1 \) nor \( \set b_2 \) can be identical to \( \set b \).
\end{proof}

 \bibliographystyle{elsarticle-num-names.bst} 
 \bibliography{references.bib}





\end{document}